\documentclass[rnote]{aa}
\def\mum{$\mu$m\,}

\usepackage{amsfonts}
\usepackage{amssymb}
\usepackage[dvips]{hyperref}
\usepackage{graphicx}
\usepackage{txfonts}
\usepackage{natbib}
\usepackage[below]{placeins}
\bibpunct{(}{)}{;}{a}{}{,} 

\begin{document}
\bibliographystyle{aa}

\title{The core flux of the brightest 10$\mu$m galaxies in the southern sky\footnote{
Based on observations collected at the European Southern
Observatory, Chile, programme numbers  67.B-0582, 68.B-0575, 68.B-0033} }

\author{D.Raban\inst{1} \and B.Heijligers\inst{1} \and
H.R\"ottgering\inst{1} \and K.Meisenheimer\inst{2} \and W.Jaffe\inst{1}
\and H.U.K{\" a}ufl\inst{3} \and T.Henning\inst{2} }

\institute{ Leiden Observatory, Leiden University, P.B.9513, Leiden, 2300 RA, the Netherlands \and Max-Planck-Institut f\"ur Astronomie,
 K\"onigstuhl 17, 69117 Heidelberg, Germany \and ESO, Karl-Schwarzschild-Str. 2, D-85748 Garching bei Muenchen, Germany}

\date{Received 3 September 2007 / Accepted 26 March 2008}

\abstract  {}
 {Near diffraction-limited images have been taken at 8.9, 11.9, and 12.9 \mum for
 the brightest extragalactic sources in the southern sky, in order to optimally plan N-band
observations with MIDI (MID-infrared Interferometric instrument) at the VLTI.  }
{We have assembled a sample of 21 objects consisting of all the AGNs
 observable from Paranal observatory, Chile, plus three non-AGN objects, with an estimated
 N-band flux greater than 400mJy. We used the TIMMI2 Mid Infrared instrument mounted on the
ESO's 3.6m telescope to obtain near diffraction-limited images in order to establish the unresolved
core flux within $<0.5$ arscsec.}
 {Positions and core total fluxes were obtained for all sources in our
 sample and compared with similar investigations in the literature.
We find that 15 AGN and the nuclear starburst in NGC 253 exhibit an unresolved core flux $<300$mJy at 11.9\mum,
making them promising targets for  MIDI at the VLTI.
For extended sources, near diffraction-limited images are presented and discussed.} {}
\authorrunning{D.Raban et al. }
\titlerunning{Brightest southern 10 $\mu$m AGNs}
\maketitle
\section{Introduction}
Since the operation of MIDI  at the VLTI  began in 2004,  astronomy has entered a new era where
 it is now possible to reach a resolution of a few milliarcseconds at infrared wavelengths, about 15 times
 the resolution of the largest single-dish telescopes. MIDI (operates in the
 N-band (8-13\mum) and therefore it is ideal for observing heated dust expected in AGNs and starburst galaxies.
   Although it is most commonly
 used in observing galactic objects like circumstellar disks \citep{Leinert04} and dust shells
around evolved stars \citep{Ohnaka05},  MIDI   has been successfully used to observe extragalactic objects.
  The main limitation of MIDI in this context is its limiting flux of $\sim$400mJy, so only bright objects can be
  observed currently.  So far, MIDI has been used to observe dusty torii in AGNs, objects which are central
 to the AGN unification model, and  which have proven too small to be resolved by a single dish telescope.
  With its superior resolution, the presence of a torus-like structure has been established in the galaxy NGC
 1068 /citep{Jaffe04} by using only two baseline observations, and additional MIDI observations with
 extensive coverage of up to 21 baselines have measured the geometrical properties of the torus in the
Circinus galaxy (\citep{Tristram07}) and NGC 1068 (Raban et at, submitted). A resolved dust structure
 was also detected in the core of Centaurus A \citep{Meisenheimer07}.
Apart from active galactic nuclei, the only  extragalactic objects bright enough for MIDI  are  starburst galaxies.
For  such objects, MIDI can be used to measure the size and geometry of the nuclear star-forming region.

Here, we have assembled a list of the brightest extragalactic objects at 10\mum, and observed them
at 8.9 ,11.9 and 12.9 \mum  with the TIMMI2 instrument, mounted on the 3.6m telescope in La Silla, Chile.
Our goal here is to give  the basic information needed in order to plan VLTI
observation of cores of galaxies.   Positions and core fluxes were determined for all sources, and for
those sources which were resolved, we present and discuss images at 8.9, 11.9 and 12.9 \mum .

 The layout of the paper is as follows: in Section \ref{sec:selection} we describe
 how the observed sources were selected; the  observations and data reduction are discussed in Sections
 \ref{sec:observations} and \ref{sec:reduction}; images are presented in   Section \ref{sec:results} along
 with a brief description of each object and additional reference information

\section{Sample selection}\label{sec:selection}

The target list consists of \emph{all} AGN's 
observable from Cerro Paranal ($\delta<22$) and have an estimated total
N-band flux density brighter than $N=5$ or $S_N=400$mJy, which is the
limiting magnitude for observations with MIDI,  estimated from previous N-band measurements
available in the
literature. This list of 21  objects (without NGC 1068) was taken from
the catalogue of Seyfert galaxies by \cite{Lipovetsky88} to which we
added additional sources from the compilations of \cite{Granato97} and
\cite{Maiolino95}, plus the  IR luminous galaxy  M83, the starburst galaxy
 NGC 253 and the famous quasar 3C\,273.
See Table \ref{tab:1} for a complete list.

 \section{Observations}\label{sec:observations}
The mid-infrared observations were carried out with the              
 TIMMI2 instrument on the ESO 3.6m telescope on La Silla,
Chile, on the nights of 6
and 7 of August 2001 and 8 to 10 February 2002. 
 TIMMI2 is a mid-infrared imager and spectrograph operating in the M
(5 \mum), N (10
\mum) and Q (20 \mum) atmospheric bandpasses. The camera is
equipped with a $320 \times 240$ pixel Si:As array and can operate at a
scale of 0.2 or 0.3 arcsec/pixel.  We used
the 0.2 arcsec/pixel scale resulting in a total field-of-view 
 of $64\arcsec\times 48\arcsec$.  For a
complete description of the instrument  refer to \cite{Reimann00} and
\cite{Kaufl03}.

For most of the  observations we selected the 1.2\mum wide
11.9\mum filter since it offered the best signal to noise ratio
for a given integration time.
A ratio of  flux over noise (per pixel) of about a 200 is
achieved by 40 minutes of integration time for a
 400mJy source. 
 Sources that were significantly brighter  were
observed for shorter periods of time.  Some targets were supplemented
with additional images with the the 8.6\,\mum filter
($\Delta\lambda=1.2\,$\mum) and the 12.9\,\mum filter
($\Delta\lambda=0.9\,$\mum).  The complete
observation log can be found in table \ref{tab:1}. 

Seeing conditions as monitored in the optical were highly
variable, ranging between 1 and 1.8 arcseconds on August 6 to 0.36 and
0.8 arcseconds on August 7 2001, while the humidity at ground level
remained around 5\%. During the February 2002 run the seeing was
constant at 1 arcsecond with a very high humidity at ground level
of 80\% $-$ 95\%.  In order to remove  the strongly variable background
radiation of the telescope and atmosphere in the mid infrared we used
the standard chopping and nodding method.  In this mode the secondary
mirror chops with a frequency of typically several Hz North-South and
the telescope is nodded approximately every 5 minutes East-West. As
all sources turned out to be compact at 10\mum, the small field
chop and nod mode was used, keeping the objects always in the
frame, while still avoiding overlapping of the images which might lead to loss of flux.
This effectively cuts the field-of-view of the chip in 4 pieces
and is therefore only possible for compact targets.  
such as extended star forming regions, the relatively small throw we used can substract these from the
 image, leading to an underestimation of the flux. 
During chopping the detector is continuously read-out with a frequency of
about 40 Hz and each 100 image pairs are subtracted, stacked and saved to disk.
The result is a series of chopped frames with both a positive and a negative image of the source, with an
exposure time of approximately 25 seconds per frame.
Subtracting 2 frames from different nod positions results in 2 positive
and 2 negative acquisitions of the object. We used a chop throw of 20
arcsec North-South and a nod of 30 arcsec East-West (except for the
observations of M83 where these values were 25 and 30 arcsec
respectively), allowing for both the chopped as well as the nodded
positions to fall onto the detector, while at the same time preventing
any overlap of the positive and negative images.

Due to technical problems with the autoguider there was a slow
position drift during the integrations. This was taken into account during
data reduction.

\section{Reduction}\label{sec:reduction}
Standard  reduction techniques were used  to convert a set
of raw frames into a final, photometrically calibrated image of the
source. Flat field correction was not applied since for all
ground-based mid-IR instruments no possibility of a reliable flatfield
correction exists up to now \citep{Starck99}, and since the
chopping and nodding method used to eliminate the thermal background
results in a relatively flat images.  For our data cosmic rays removal is not required since the
number of background photons is large, making the cosmic ray's
contribution undetectable.  Bad pixels were detected using the
statistical variation of each pixel, extrapolated and excluded from
the measurements.  Three different
modes for referencing and co-adding were implemented to accommodate
different SNR levels. For all these modes the image peak was
determined by fitting a Gaussian of appropriate size to the central
airy disk:
\begin{itemize}
\item{For the brightest sources with a flux above $1200$ mJy (i.e.
  Circinus, NGC1068, NGC253, NGC5128) the peak SNR was high enough so
  that each of the four images (2 positive, 2 negative) in a single frame
  could be separately fitted, referenced and co-added in order to get the
  best registration, thus compensating for possible small errors in the
  chopping distance.}
\item{For the intermediate sources with a core flux above $650$ mJy we assumed the
  chopping  distance to be fixed, first combining the positive and negative
  images before their center point is determined.}
\item{For weak sources below 650 mJy the image pairs are first co-added
  with a fixed chopping distance. Images taken just before and
  after each image are also co-added before being fitted. This procedure first averages a subset
of images, calculates the peak position and then uses this position for
next subset of images, iterating  until the corrections
are smaller then 0.1 pixel.}
\end{itemize}
Since the transmissivity of the atmosphere is
highly variable in the observed wavelengths, absolute calibration is challenging.
 The system is photometrically calibrated using bright stars of known flux that have been observed every two hours. Care was taken
  that the stars
are observed at similar altitudes to the science objects.  For the
photometry measurement a distinction is made between objects where the
nucleus is a mere point source and objects displaying extended emission.
 For point sources a $1.6\arcsec$ aperture flux was taken, covering
the central airy disk out to the first minimum. For extended objects
the peak value was used since the PSF is well sampled with $0.2\arcsec$ per
pixel. Using the peak value minimizes contamination from
 extended components. For these sources the growth curve of the source and the PSF are also presented
in order to make clear which part of the flux comes from an unresolved source. The peak flux
corresponds to an aperture of 1.2$\arcsec$.
An exception is NGC 7582, for which the flux was measured with a $2.6\arcsec$ aperture.
In order to improve the visual appearance all images have 
also been deconvolved, using a simple but robust, CLEAN method, where a PSF 
taken from a nearby reference star is iteratively subtracted from the peak 
position and replaced with a Gaussian with a similar width.
Flux errors are dominated by the calibration errors from the reference
stars measured by estimating the fluctuations of the transmissivity of the sky by comparing the fluxes of the same objects (stars or
 AGNs) close in
time. Accordingly, the photometric error is about 15\% for all sources.

\section{Results}\label{sec:results}

\begin{figure*}[htbp]
  \includegraphics[width=6cm]{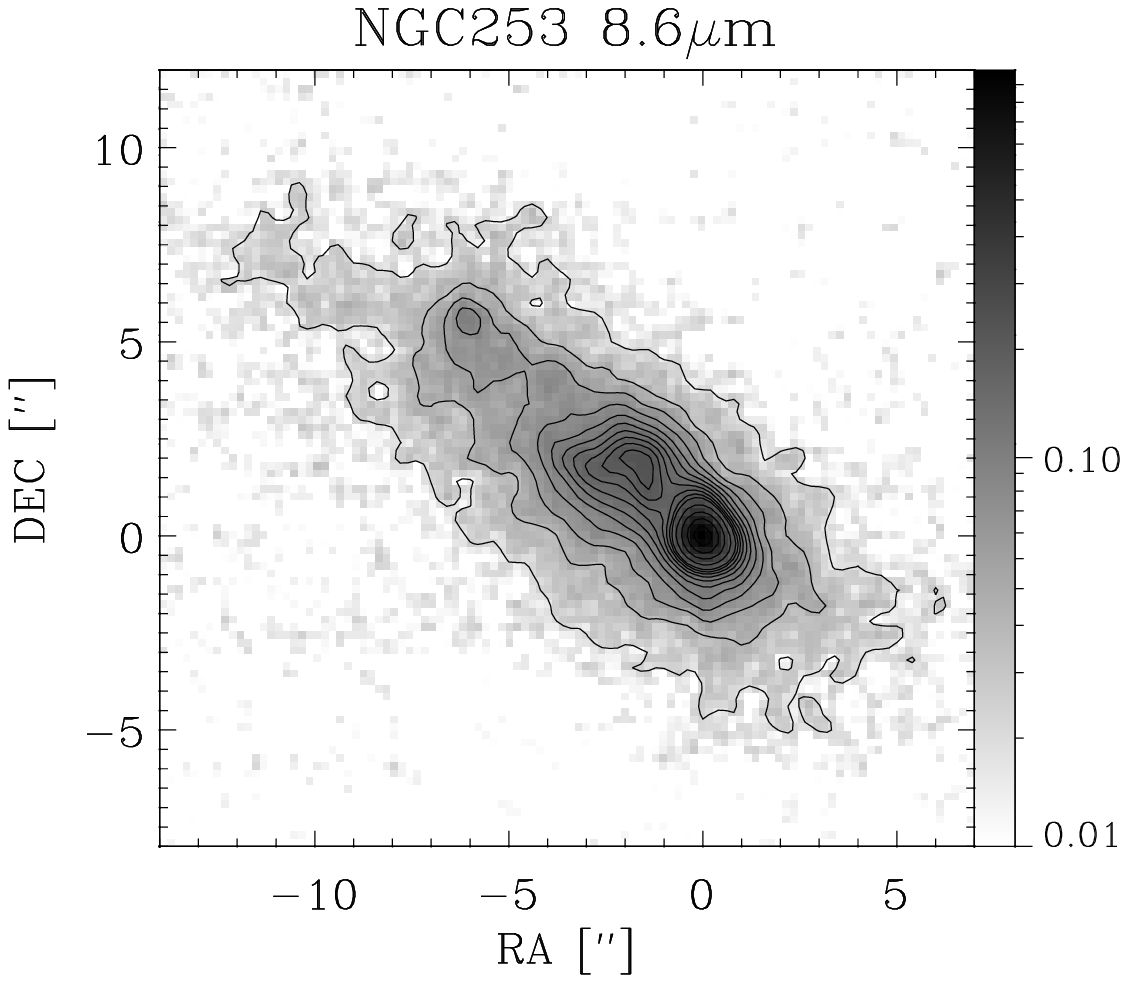}
  \includegraphics[width=6cm]{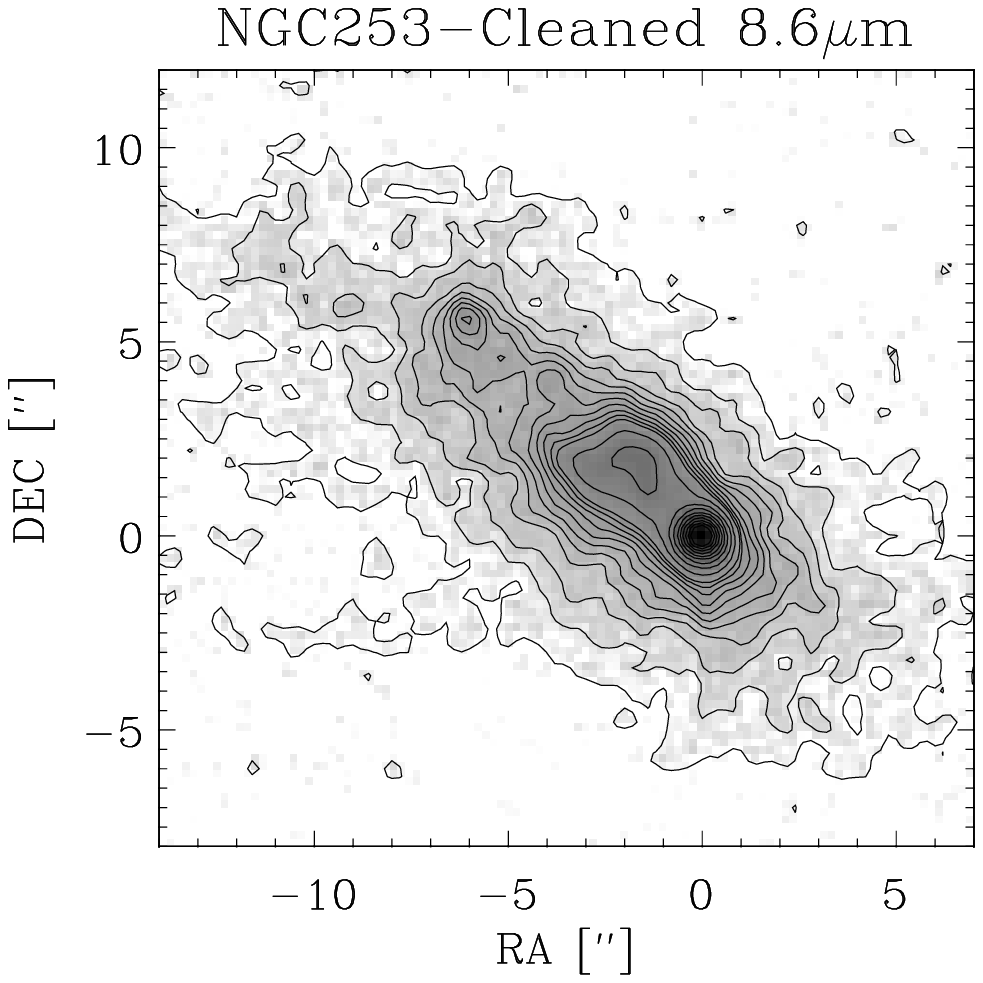}
  \includegraphics[width=6cm]{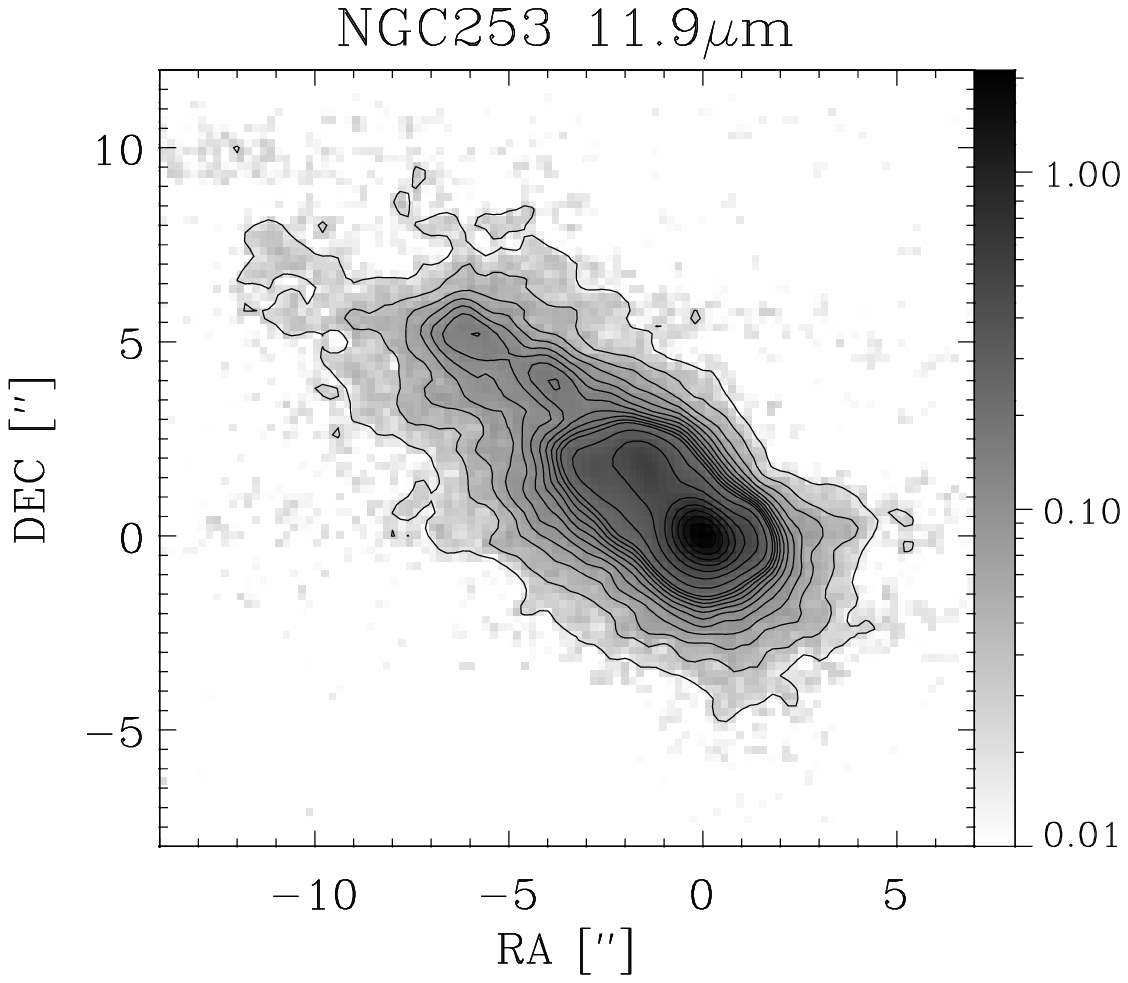}
  \includegraphics[width=6cm]{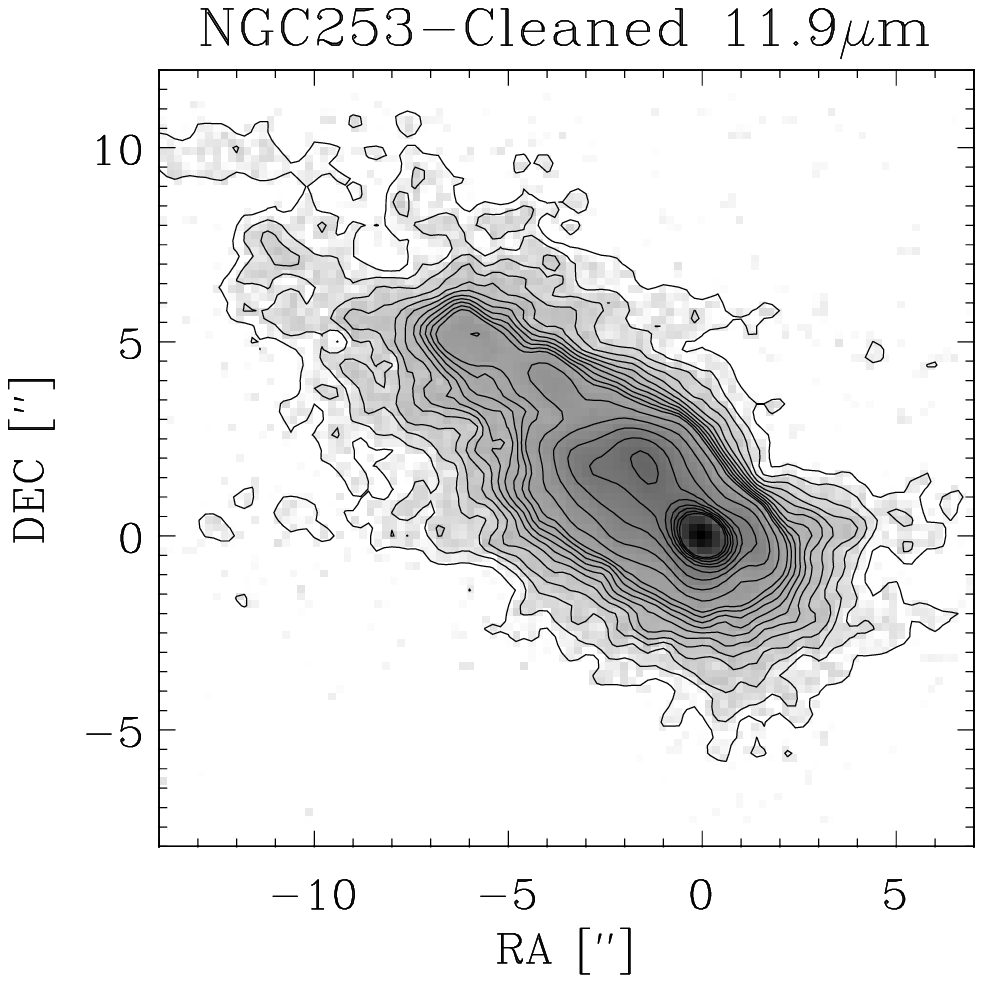}
  \includegraphics[width=6cm]{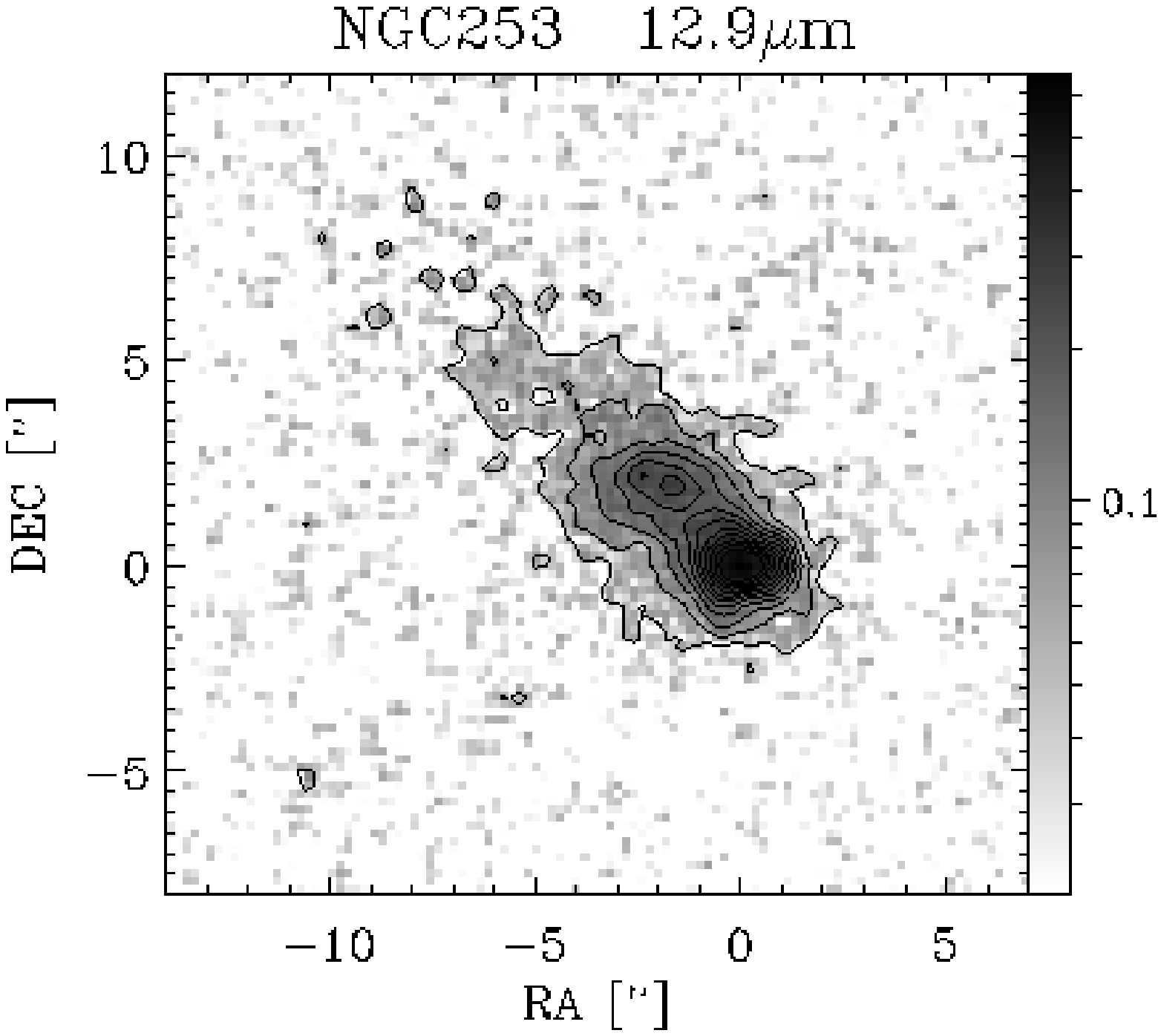}
  \includegraphics[width=6cm]{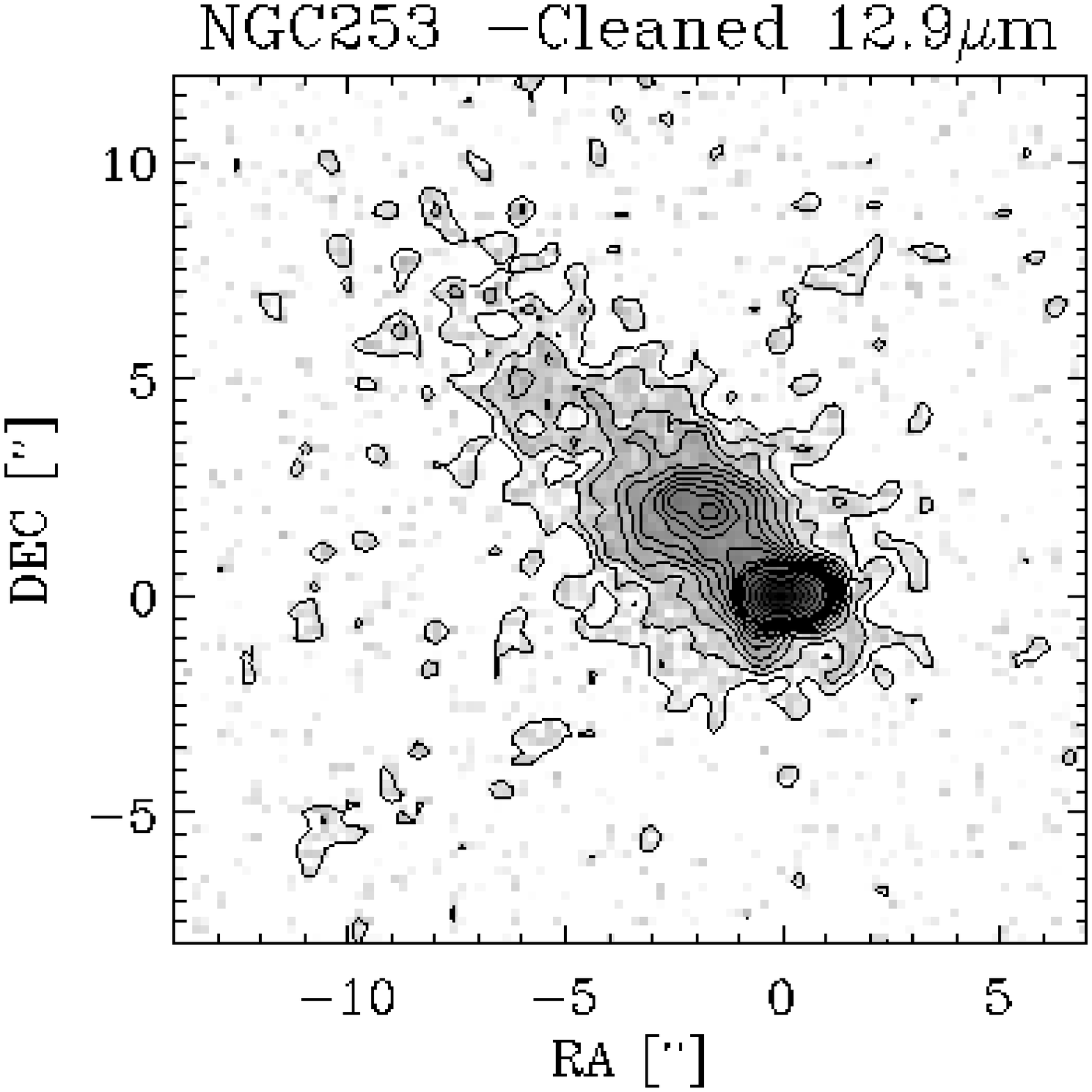}
 \includegraphics[width=6cm]{7444fg1G.eps}
\includegraphics[width=6cm]{7444fg1H.eps}
  \caption{NGC 253 contour and image overlays, Left:
    Raw data. Right: Central point deconvolved with CLEAN algorithm
sdf. Bottom : growth curve of the core and PSF at 11.9\mum (left)
and 8.9\mum (right). \label{N253b}}
\end{figure*}

\begin{figure*}[htbp]
  \includegraphics[width=8cm]{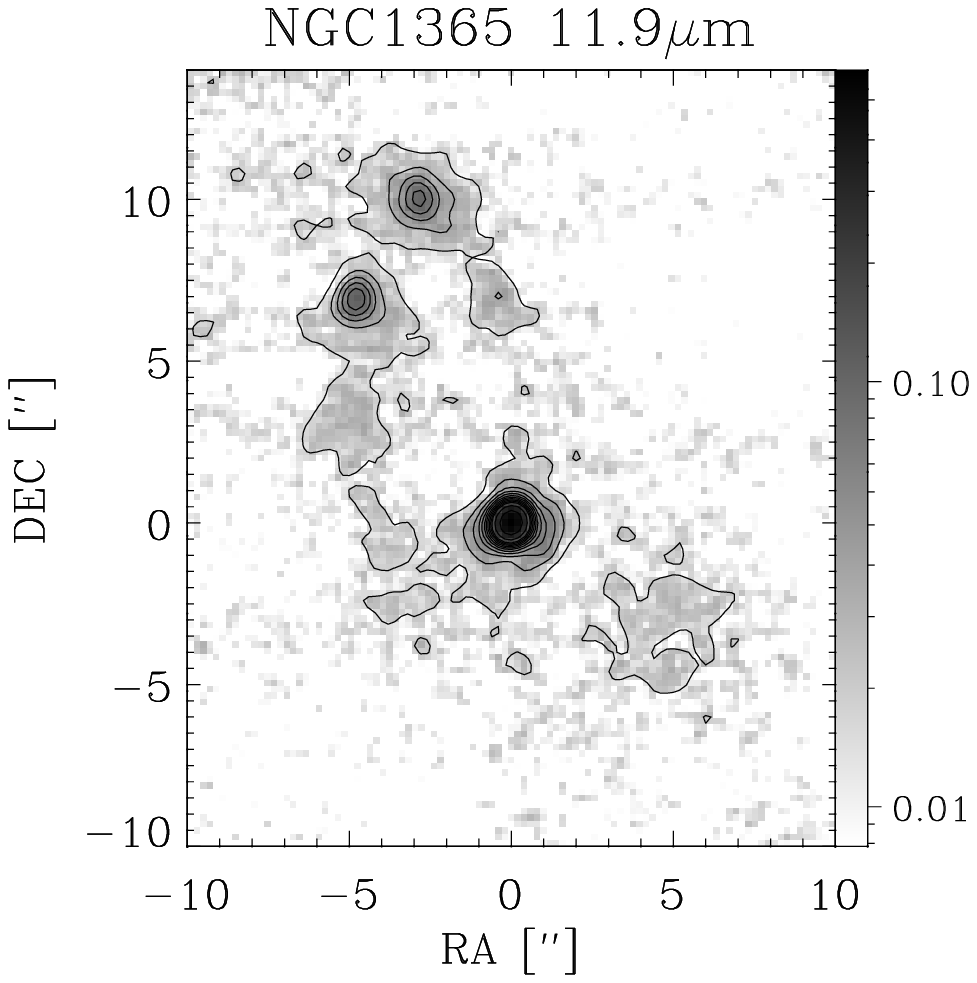}
  \includegraphics[width=8cm]{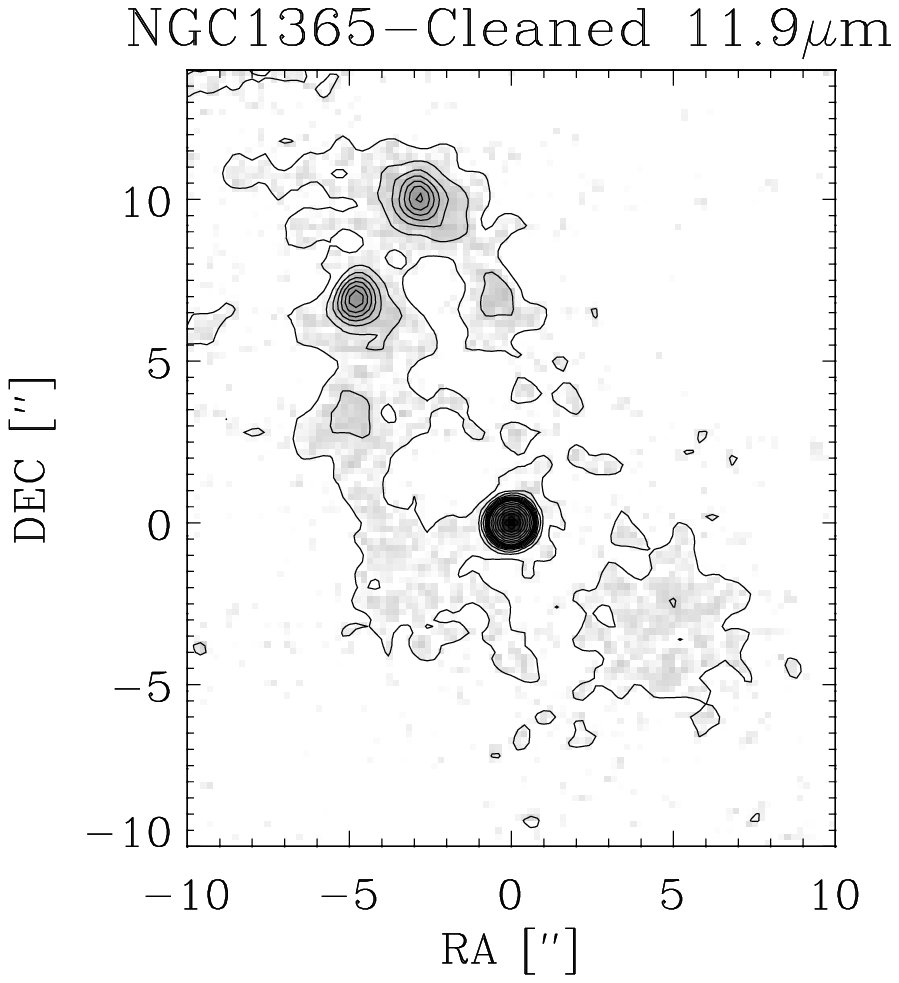}
  \includegraphics[width=8cm]{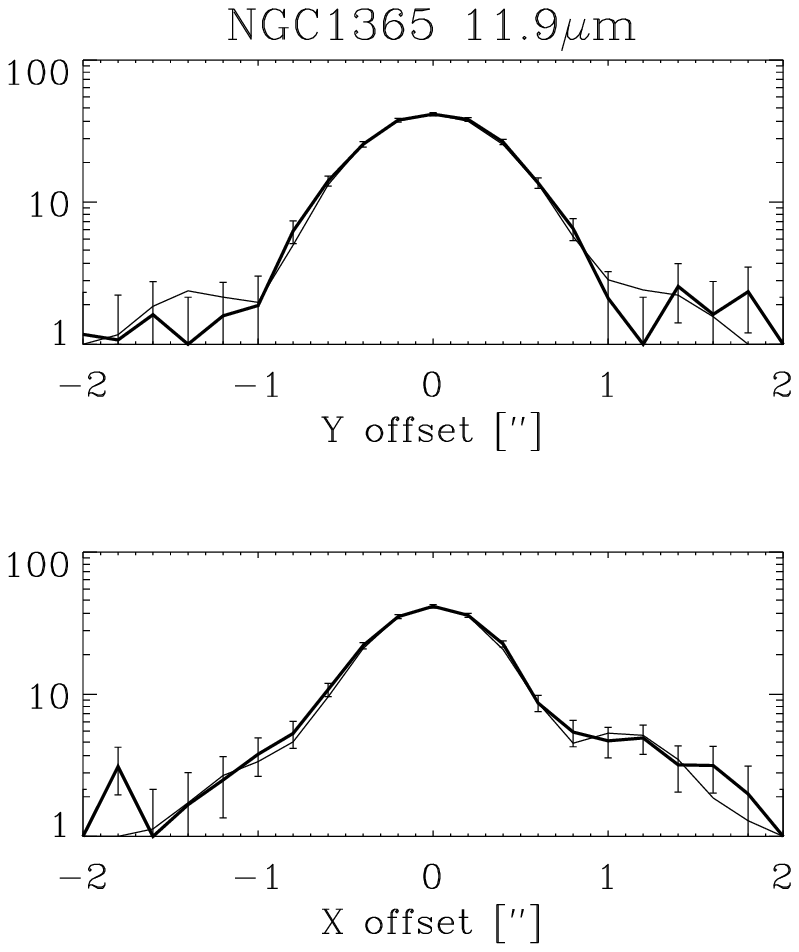}

  \caption{NGC 1365 Contour and image overlays. Left: 
    Raw data. Right: Central point deconvolved with CLEAN algorithm
sdf. Bottom: comparison of X and Y axis crosssection with psf, in normalized counts.  
\label{ngc1365}}
\end{figure*}

\begin{figure*}[htbp]
\centering 
  \includegraphics[width=8cm]{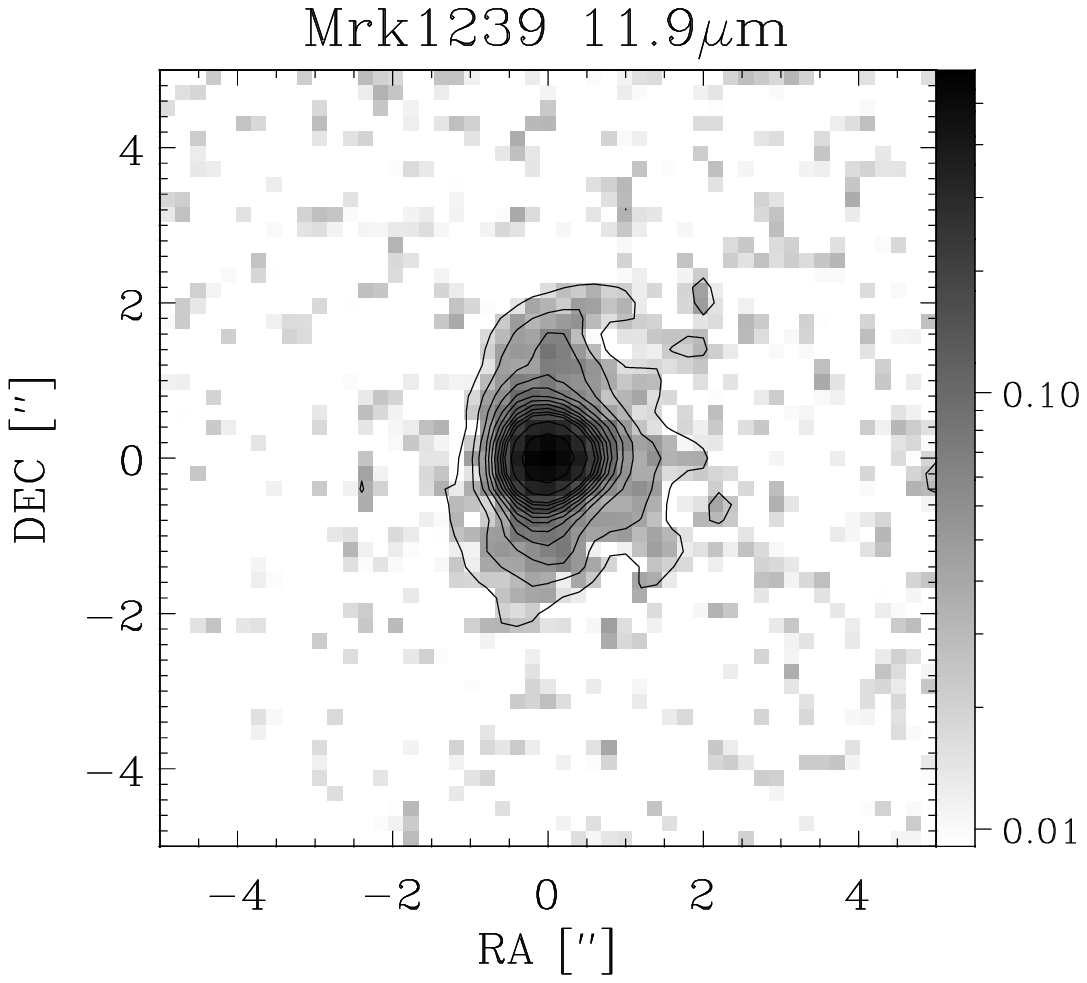}
  \includegraphics[width=8cm]{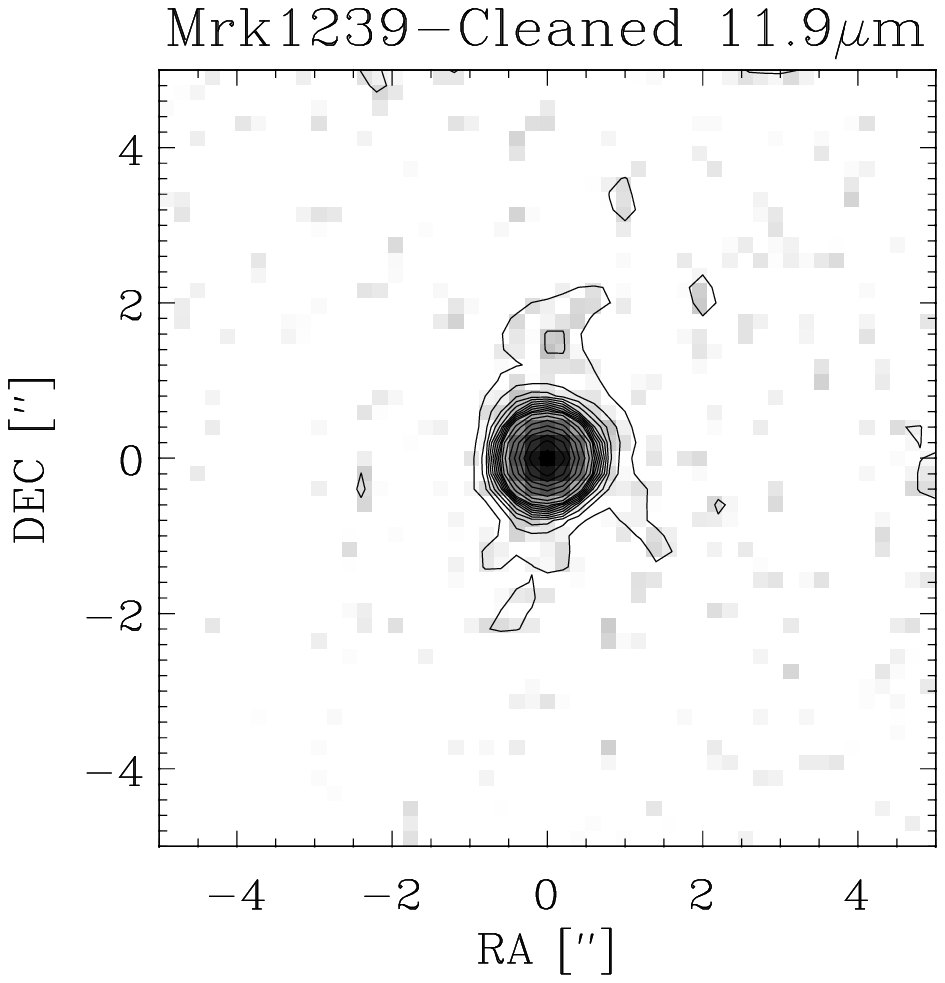}
  \includegraphics[width=8cm]{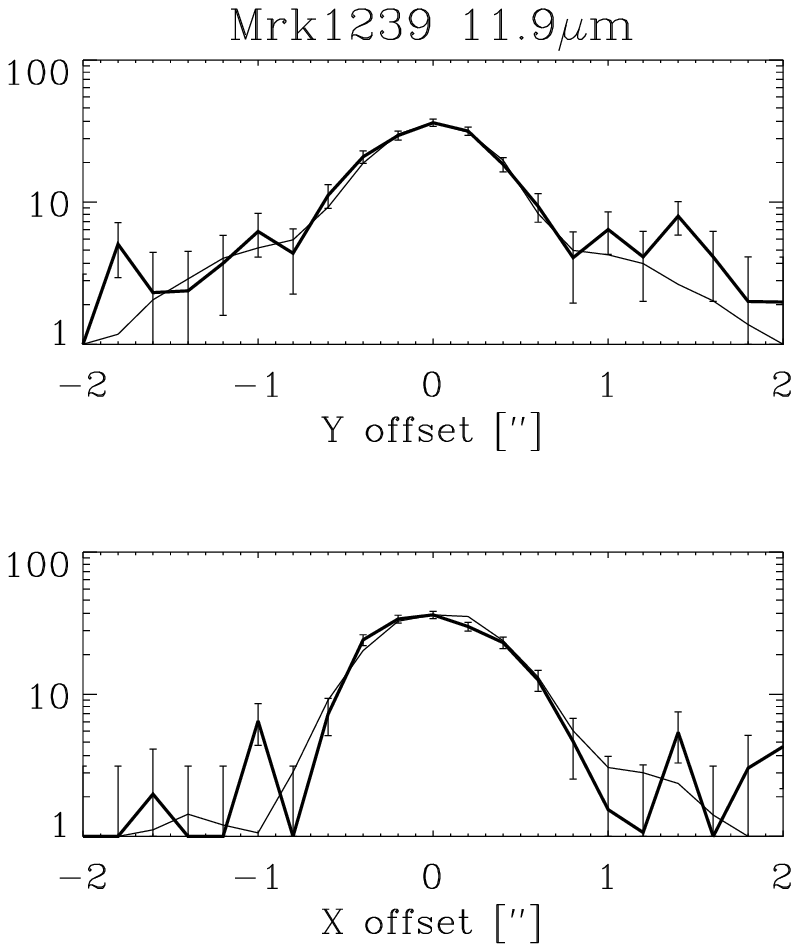}
  \caption{Mrk 1239 contour and image overlays, see Figure
\ref{ngc1365}. \label{Mrk1239}}
\end{figure*}

\begin{figure*}[htbp]
\includegraphics[width=7cm]{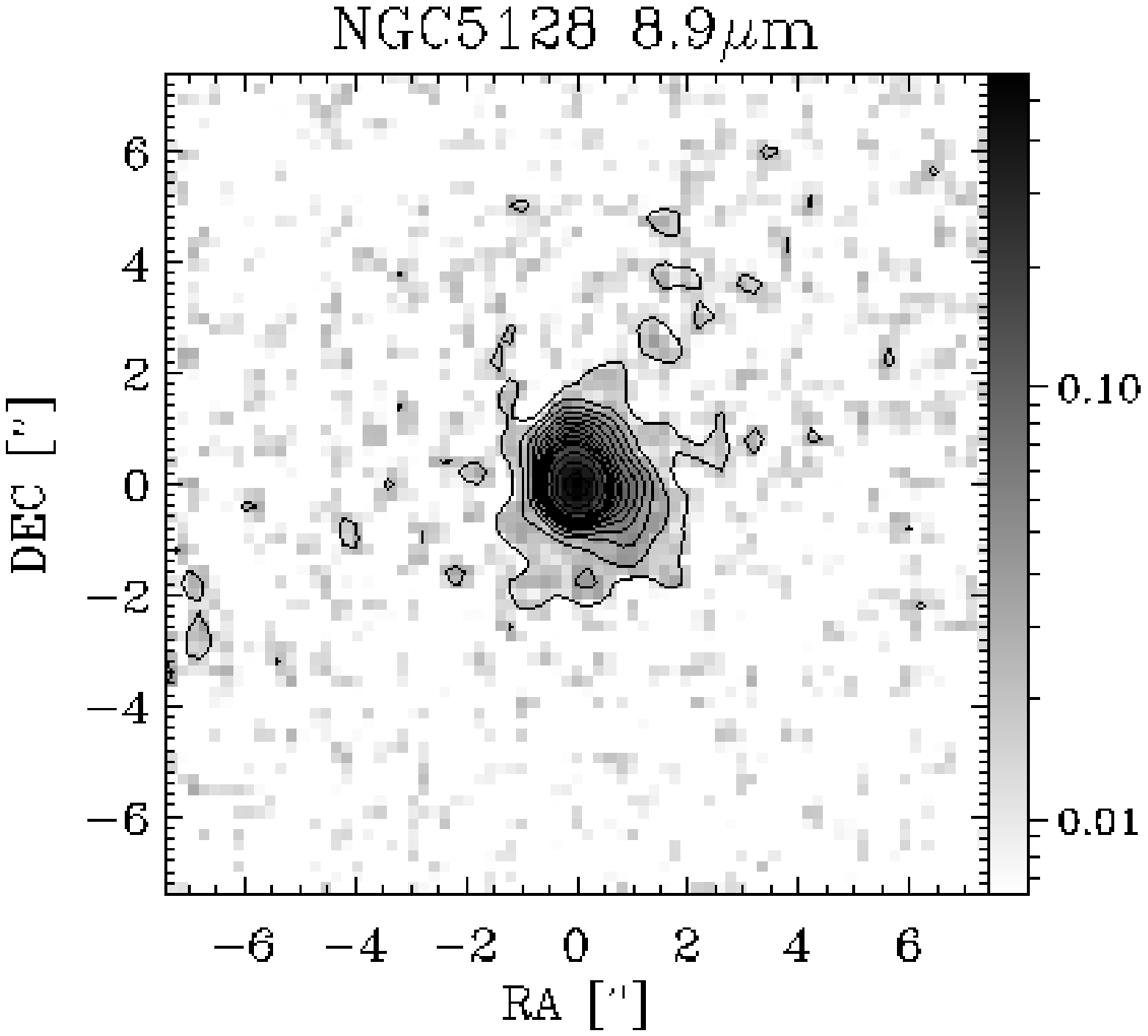}
  \includegraphics[width=7cm]{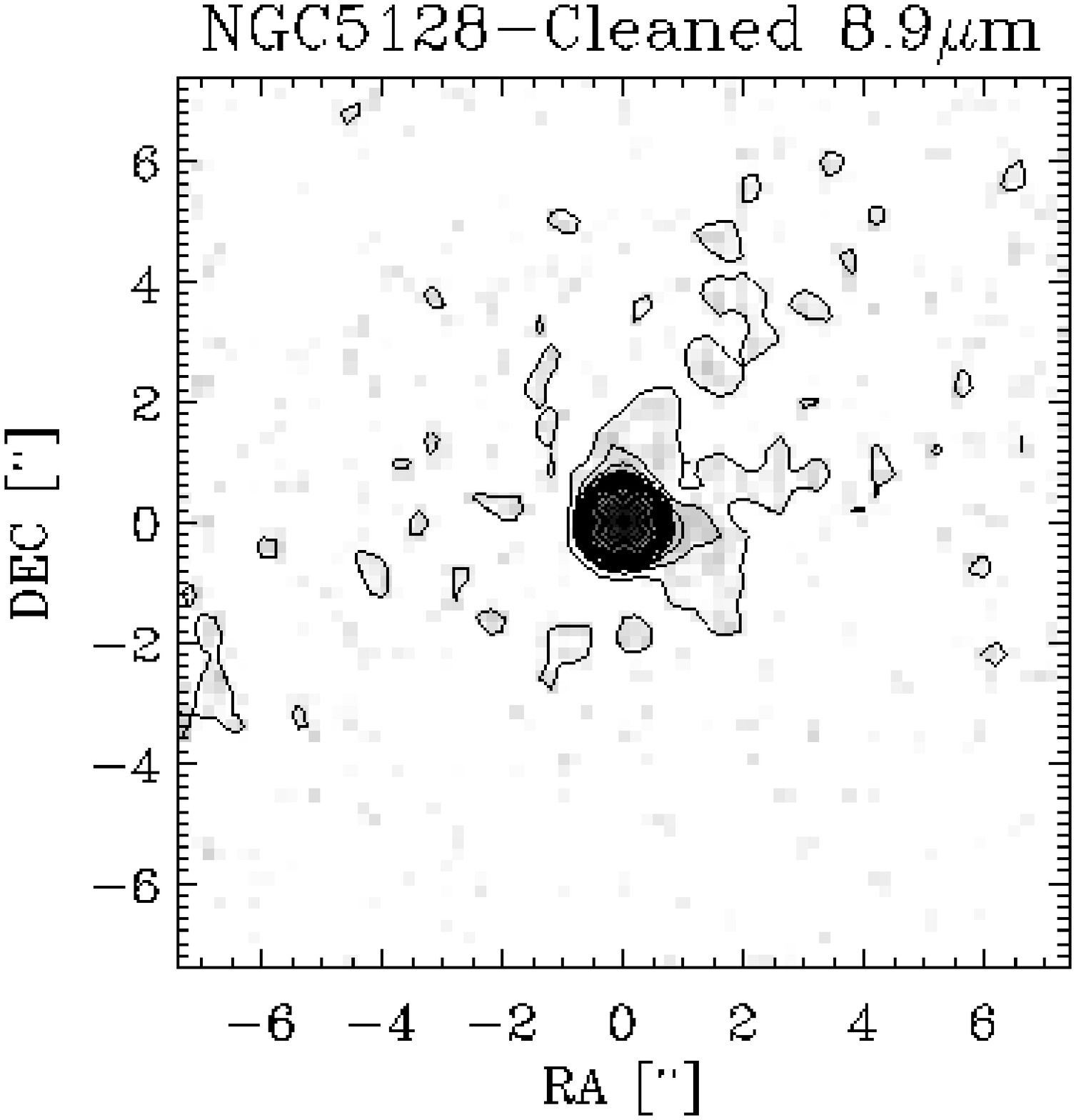}
 \includegraphics[width=7cm]{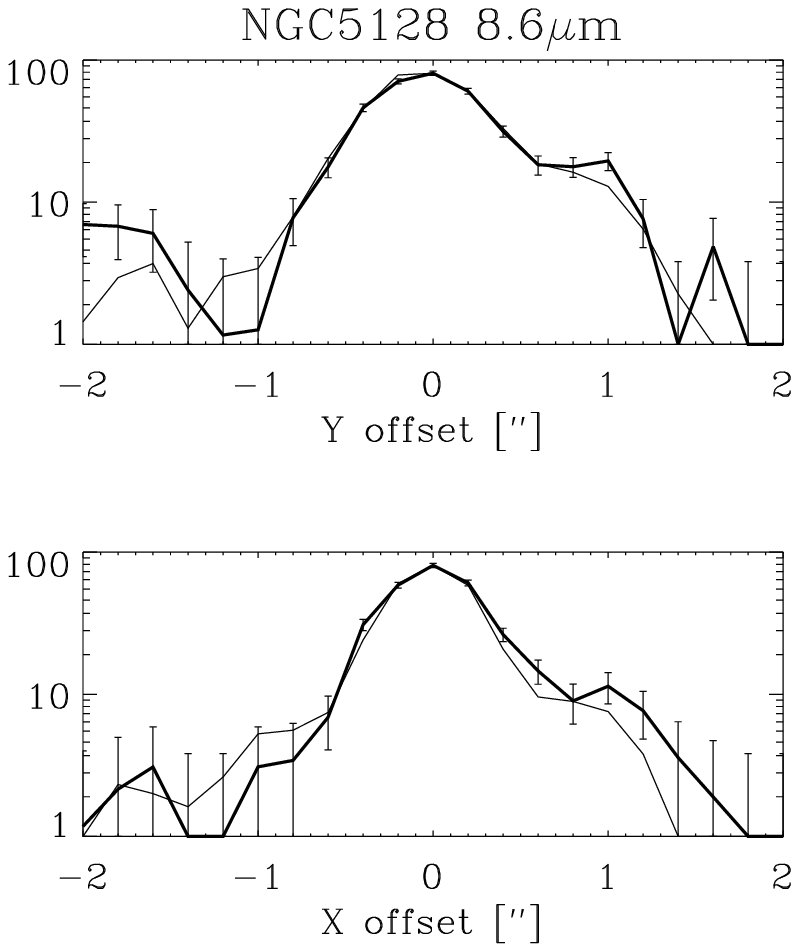}
  \includegraphics[width=7cm]{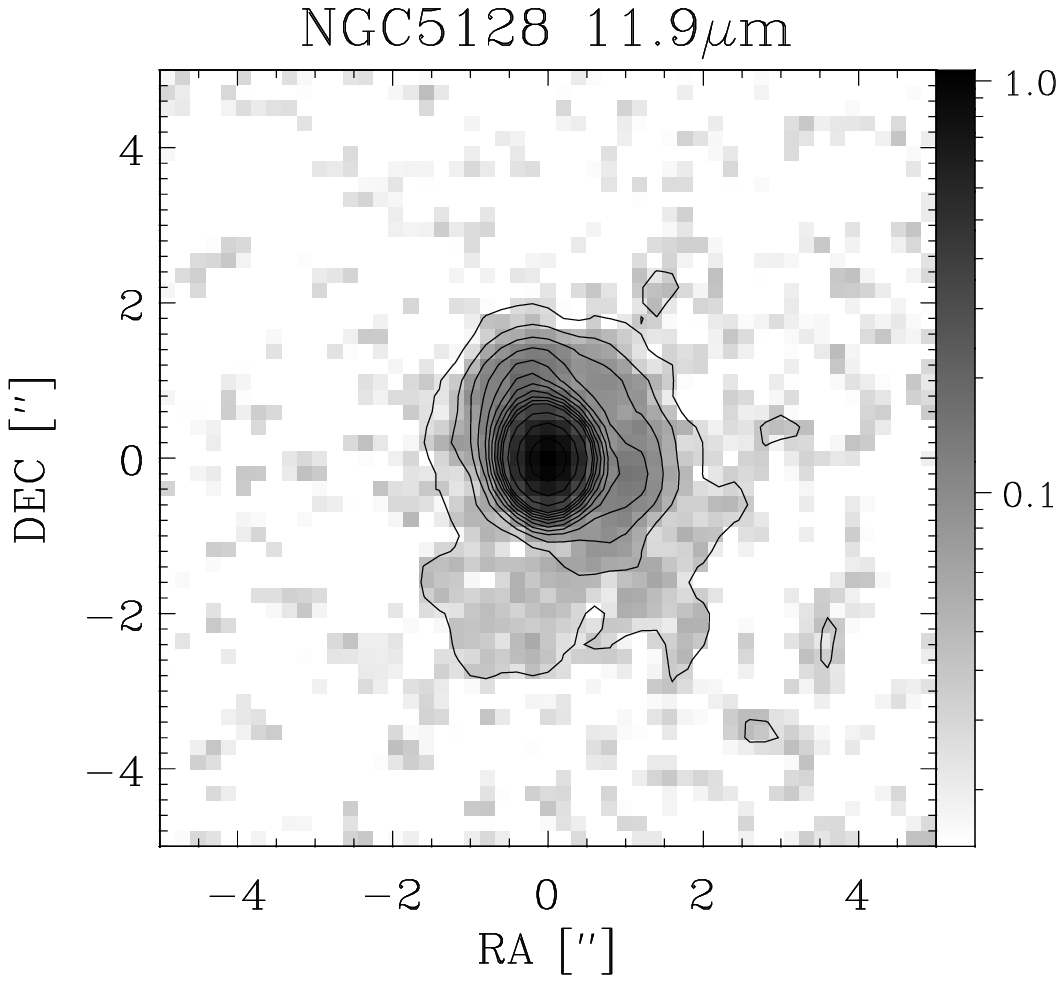}
  \includegraphics[width=7cm]{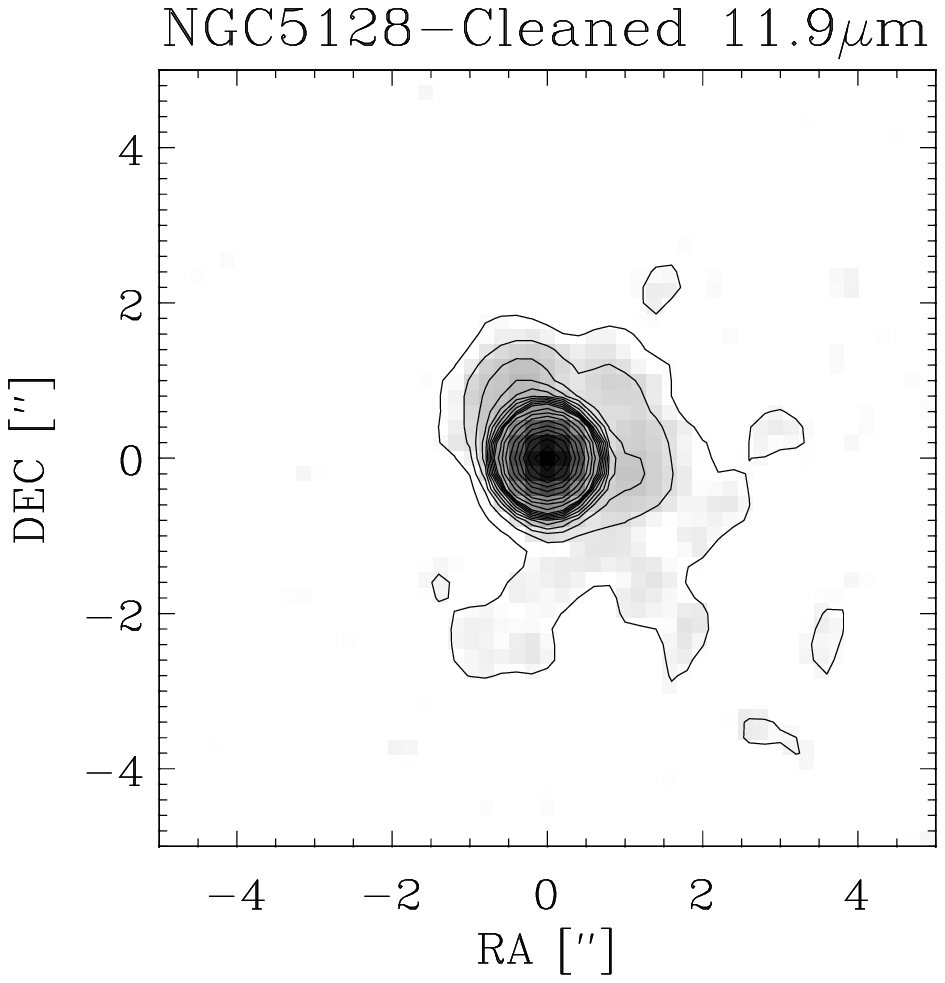}
 \includegraphics[width=7cm]{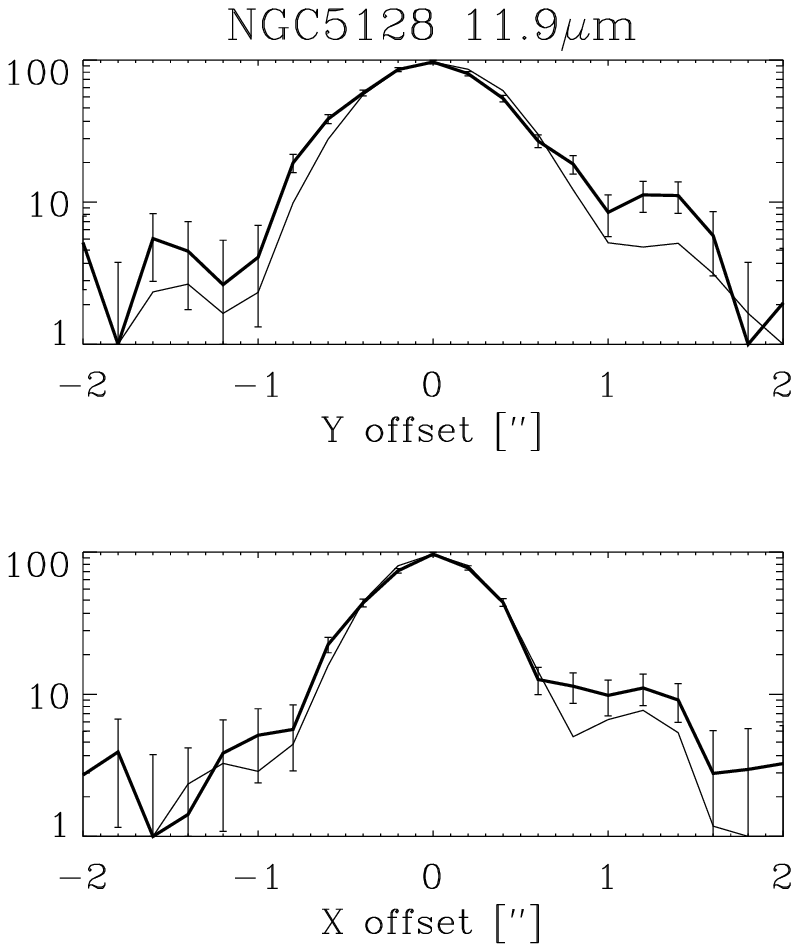}

  \caption{NGC 5128 contour and image overlays, see Figure
\ref{ngc1365}. \label{N5128}}
\end{figure*}

\begin{figure*}[htbp]
  \includegraphics[width=8cm]{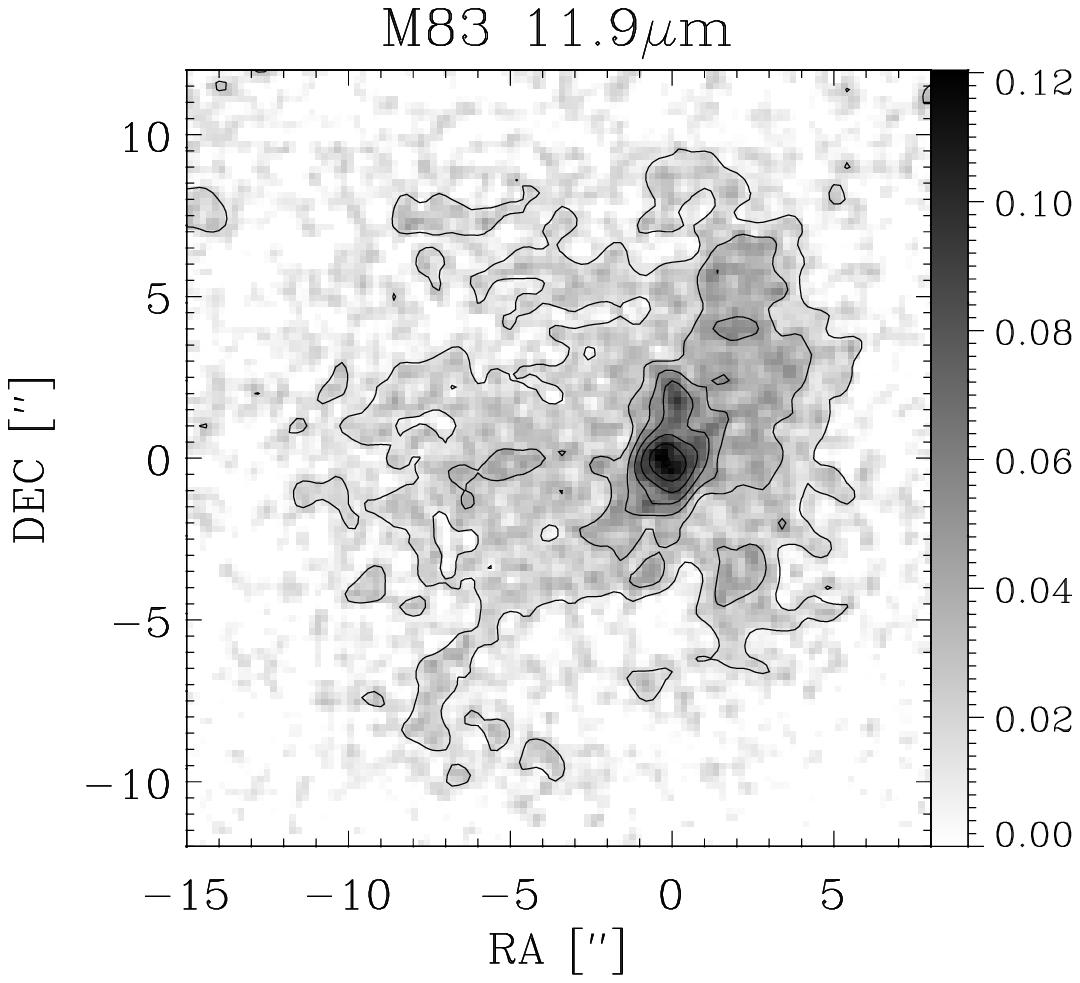}
  \includegraphics[width=8cm]{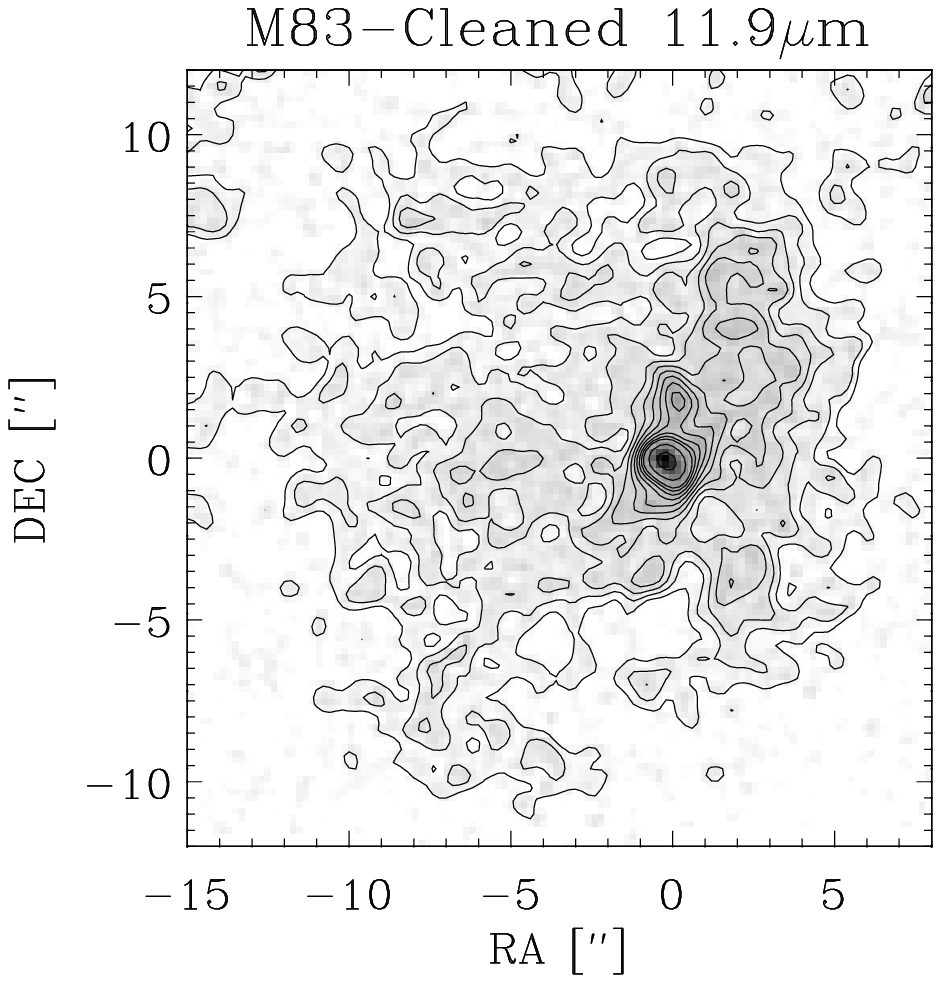}
 \includegraphics[width=8cm]{7444fg5C.eps}
  \caption{M 83 contour and image overlays, see Figure \ref{ngc1365}.
   Bottom: growth curve of core compared with the PSF.
\label{M83}}
\end{figure*}

\begin{figure*}
  \includegraphics[width=7cm]{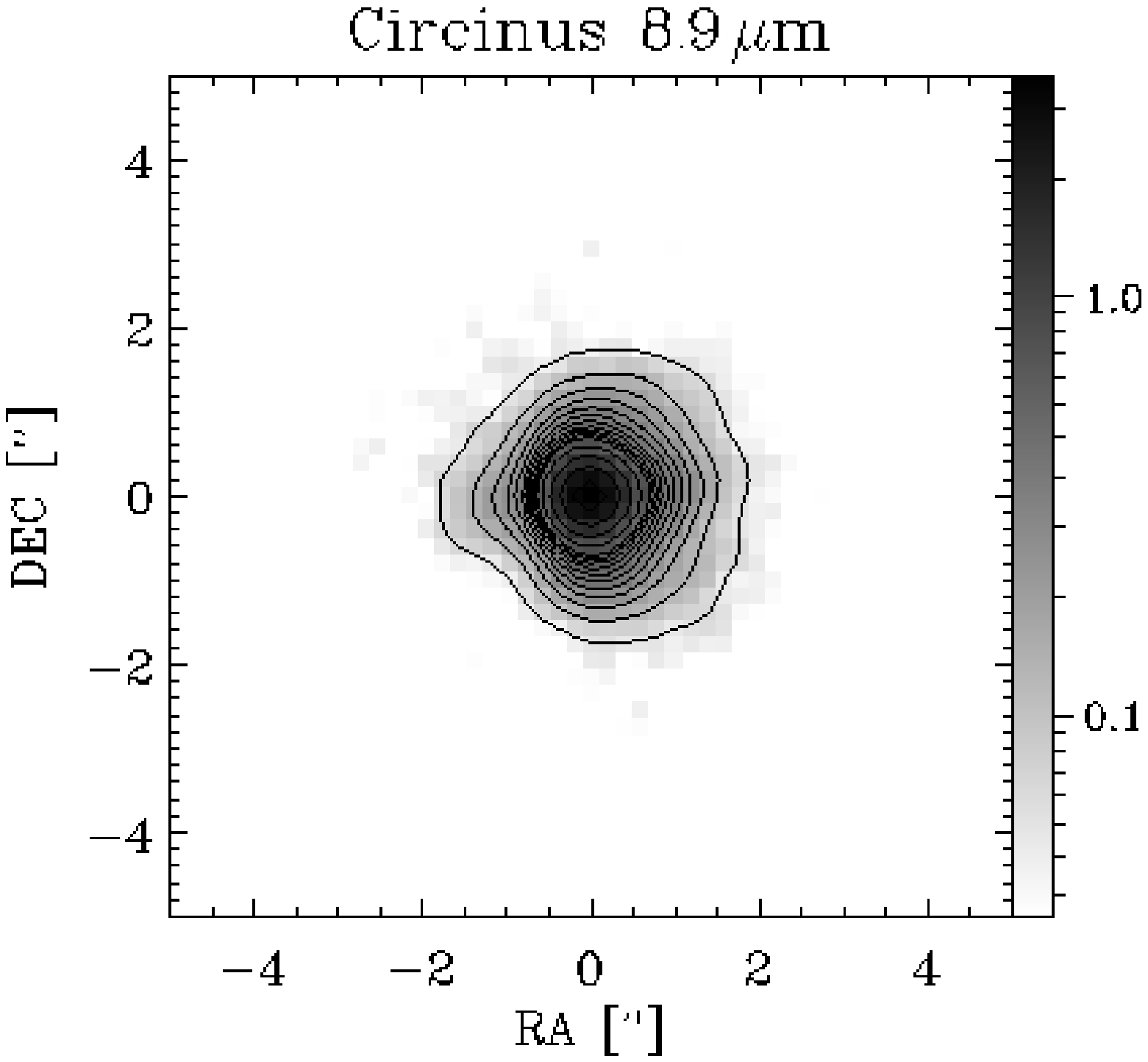}
  \includegraphics[width=7cm]{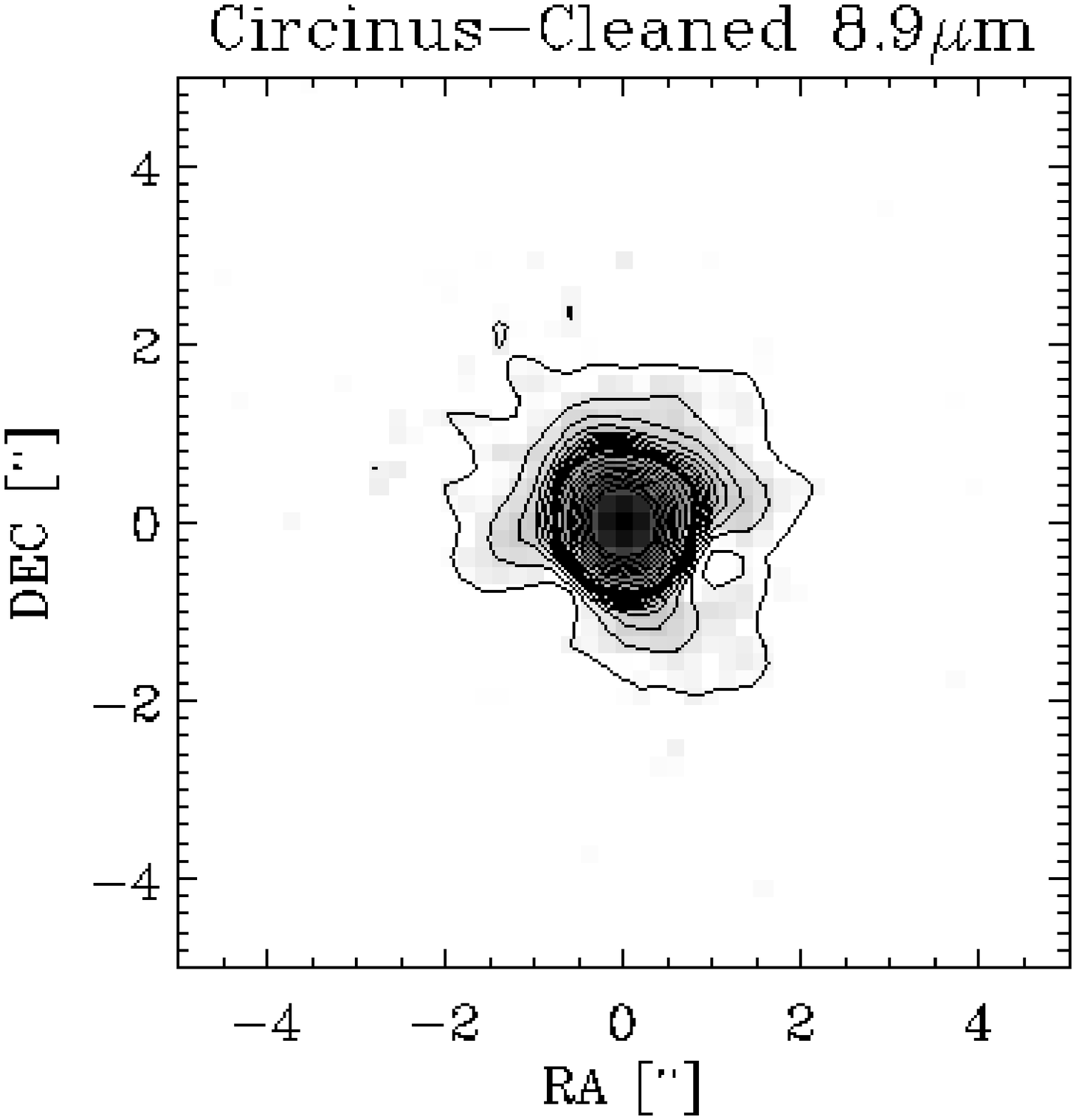}
  \includegraphics[width=7cm]{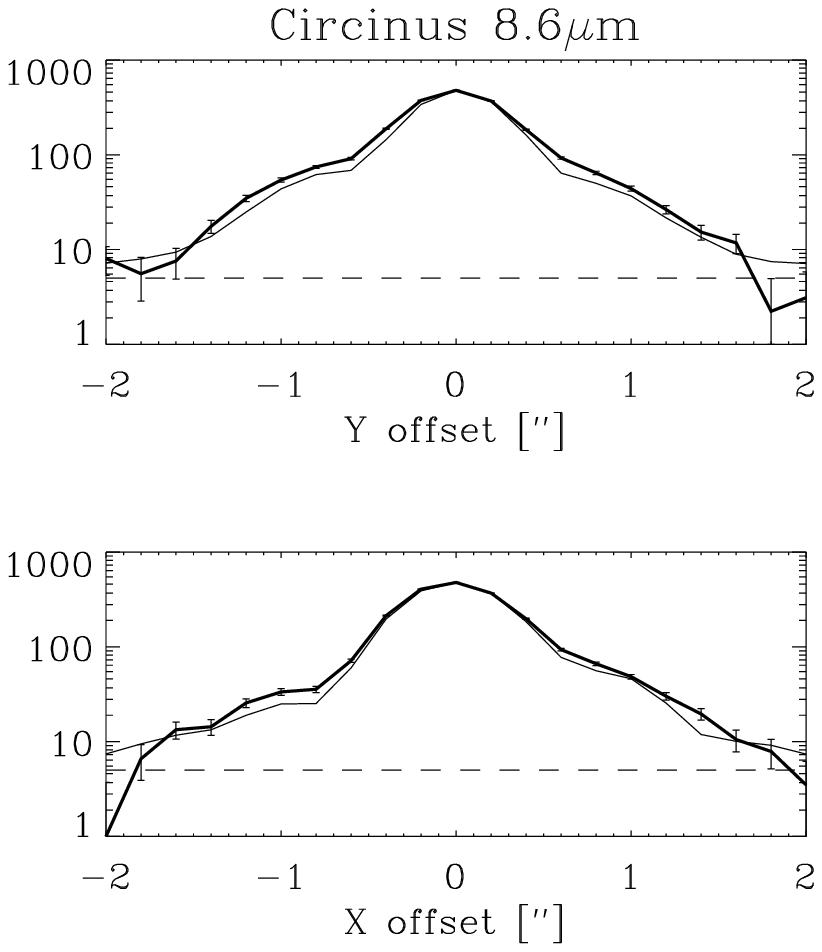}
  \includegraphics[width=7cm]{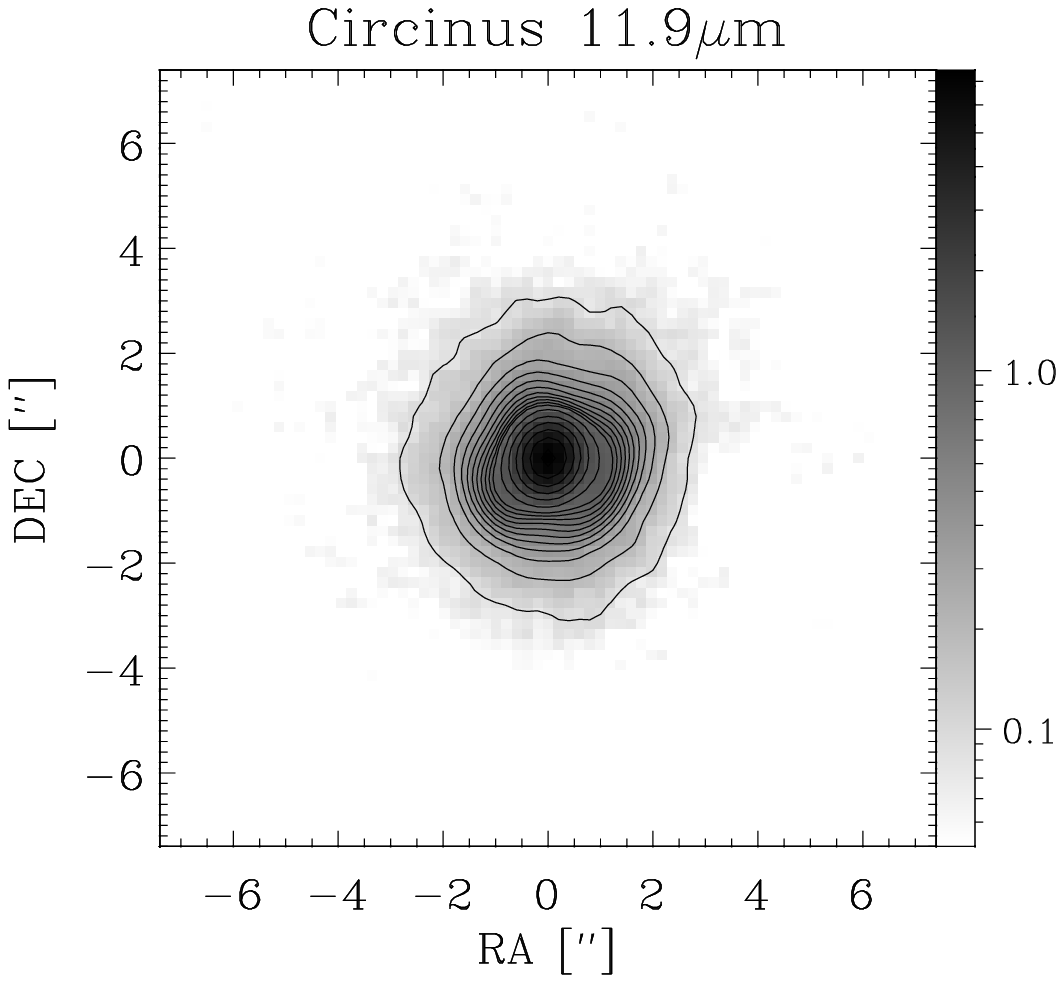}
  \includegraphics[width=7cm]{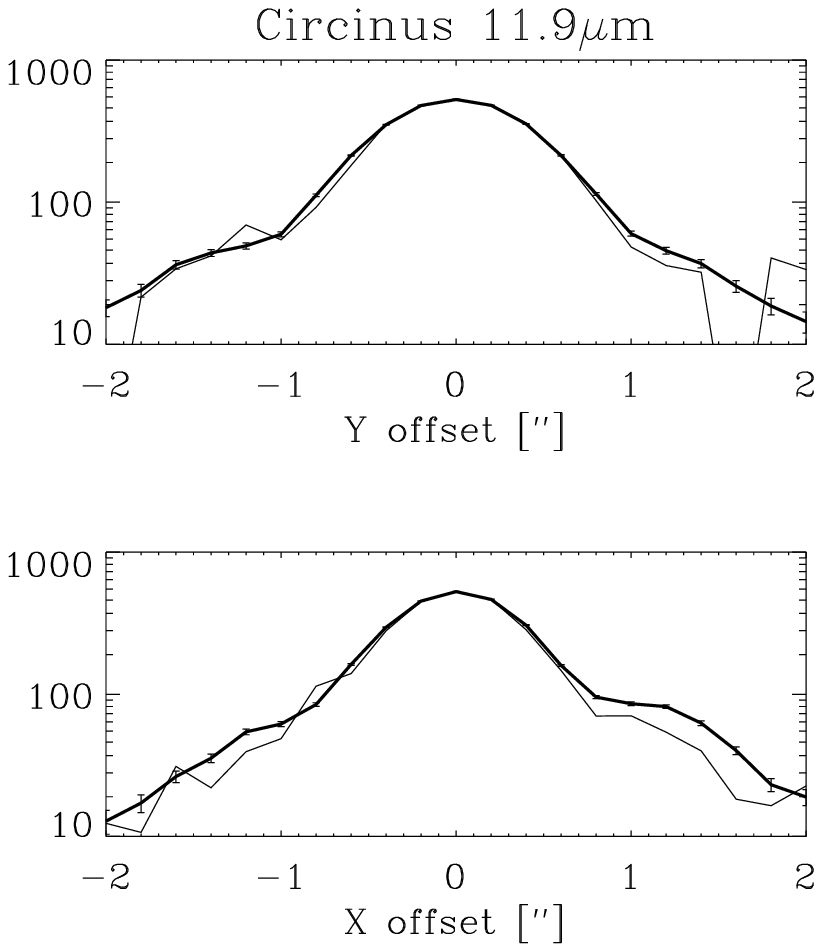}
  \caption{Circinus contour and image overlays, see Figure
\ref{ngc1365}. \label{circinus}}
\end{figure*}

\begin{figure*}[htbp]
 \includegraphics[width=7cm]{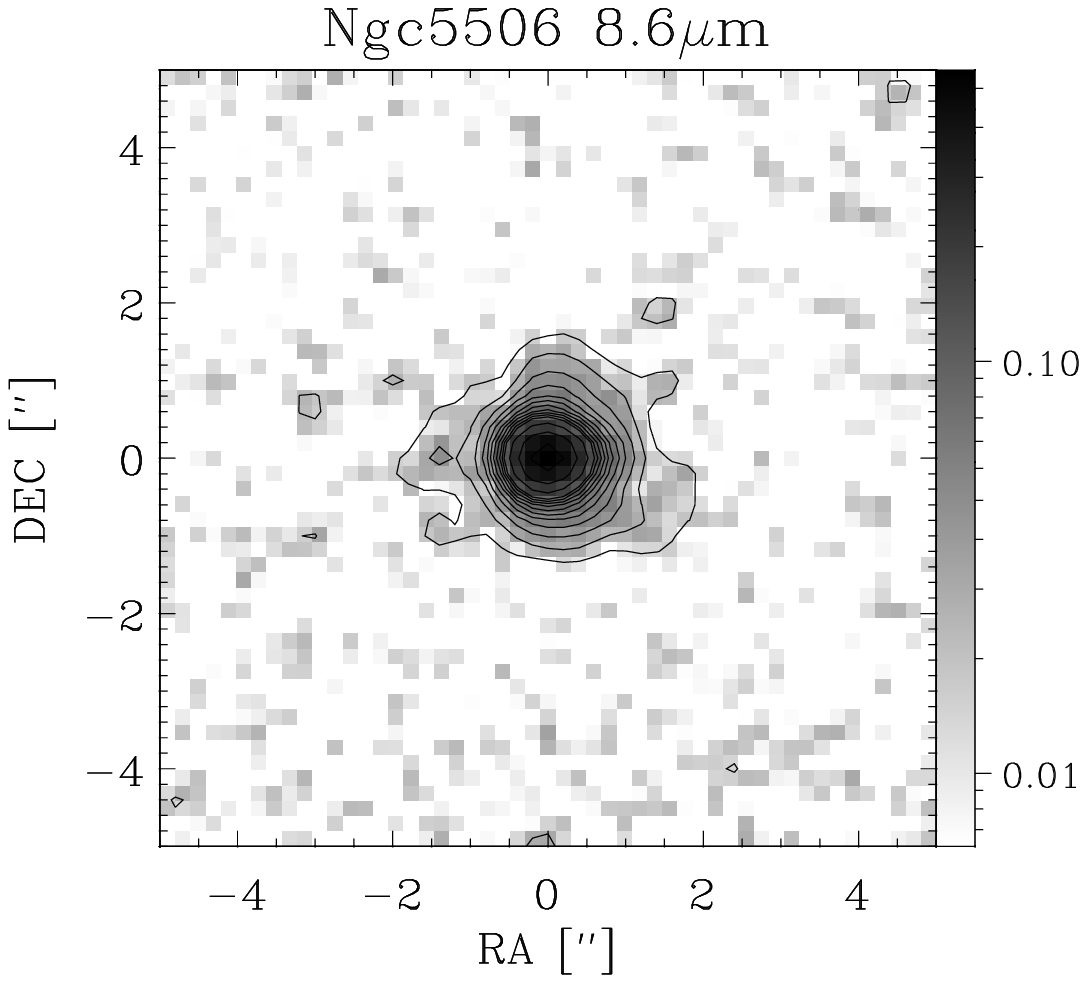}
  \includegraphics[width=7cm]{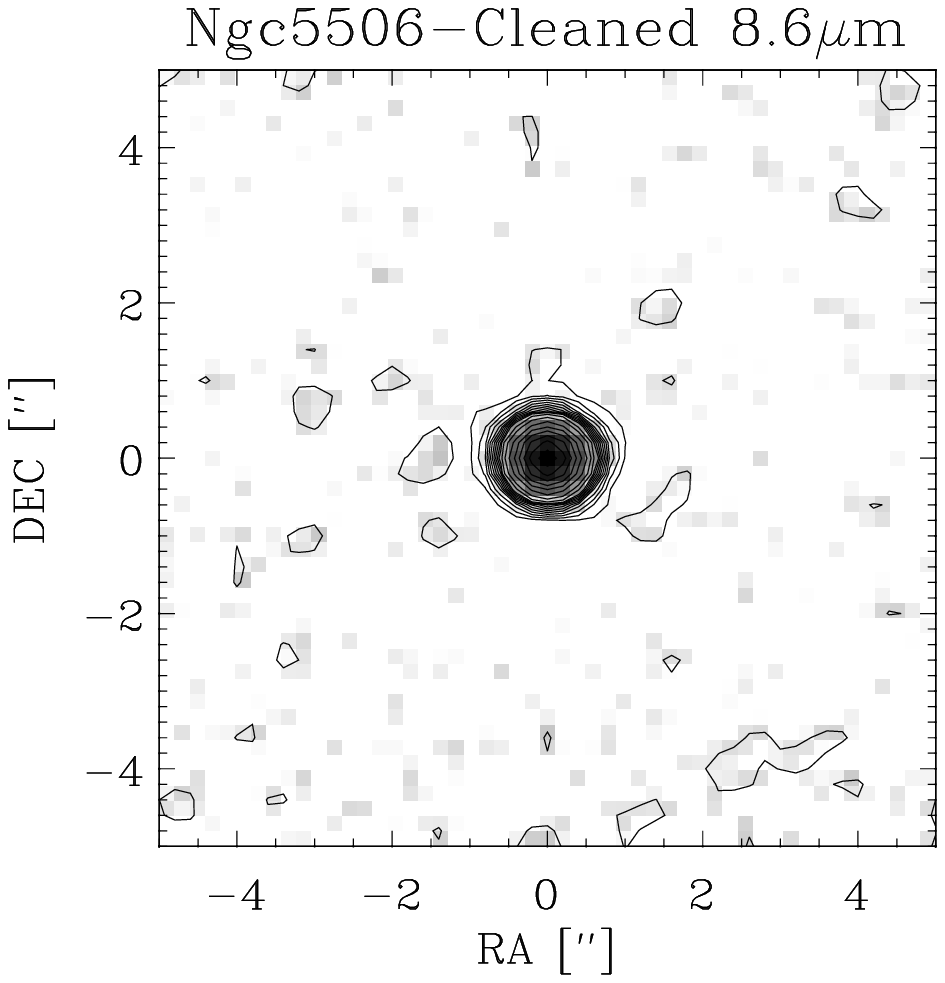}
 \includegraphics[width=7cm]{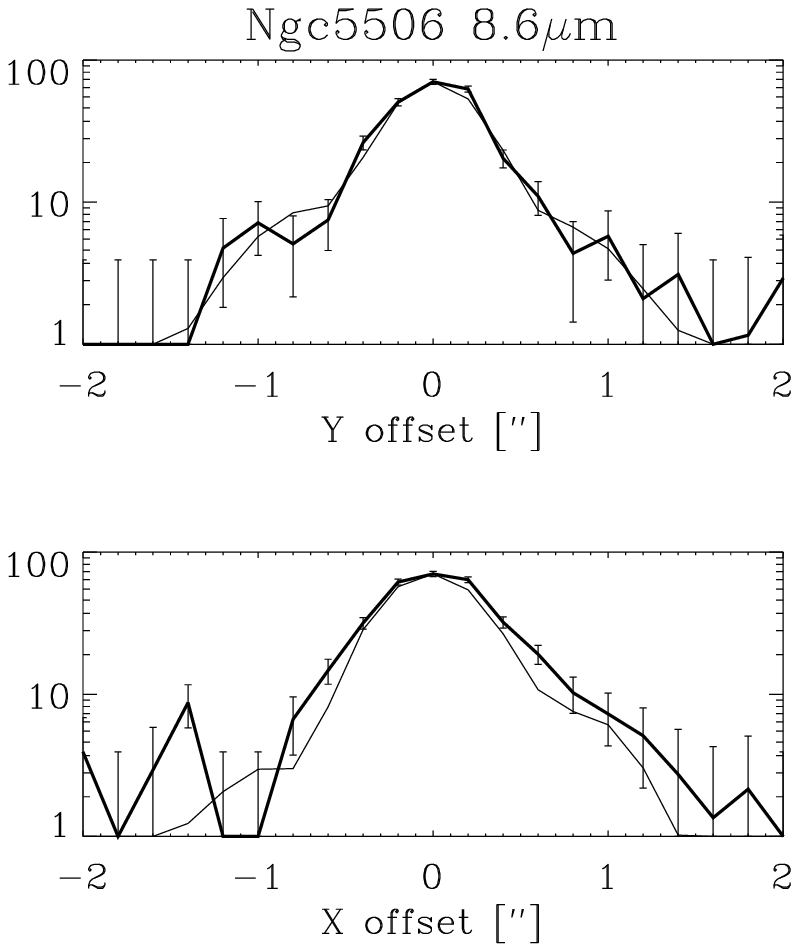}
  \includegraphics[width=7cm]{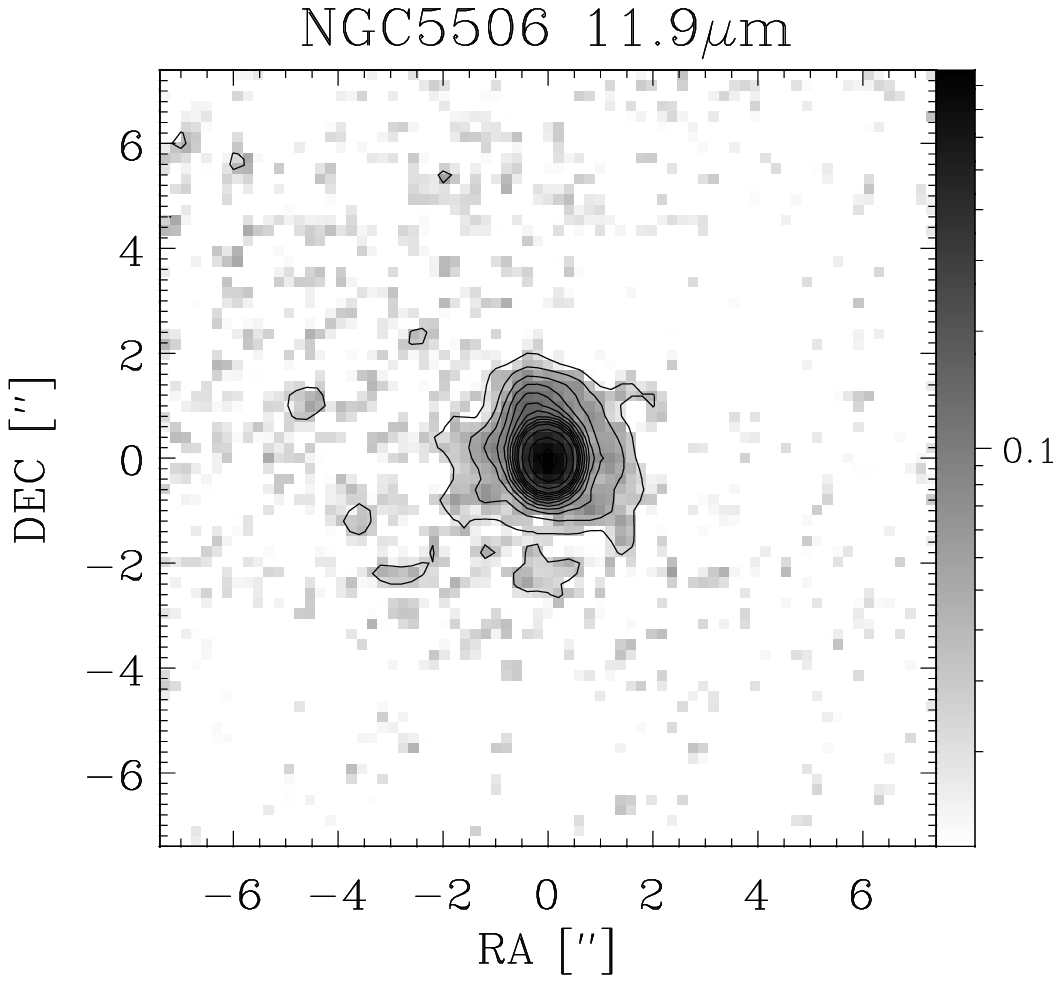}
  \includegraphics[width=7cm]{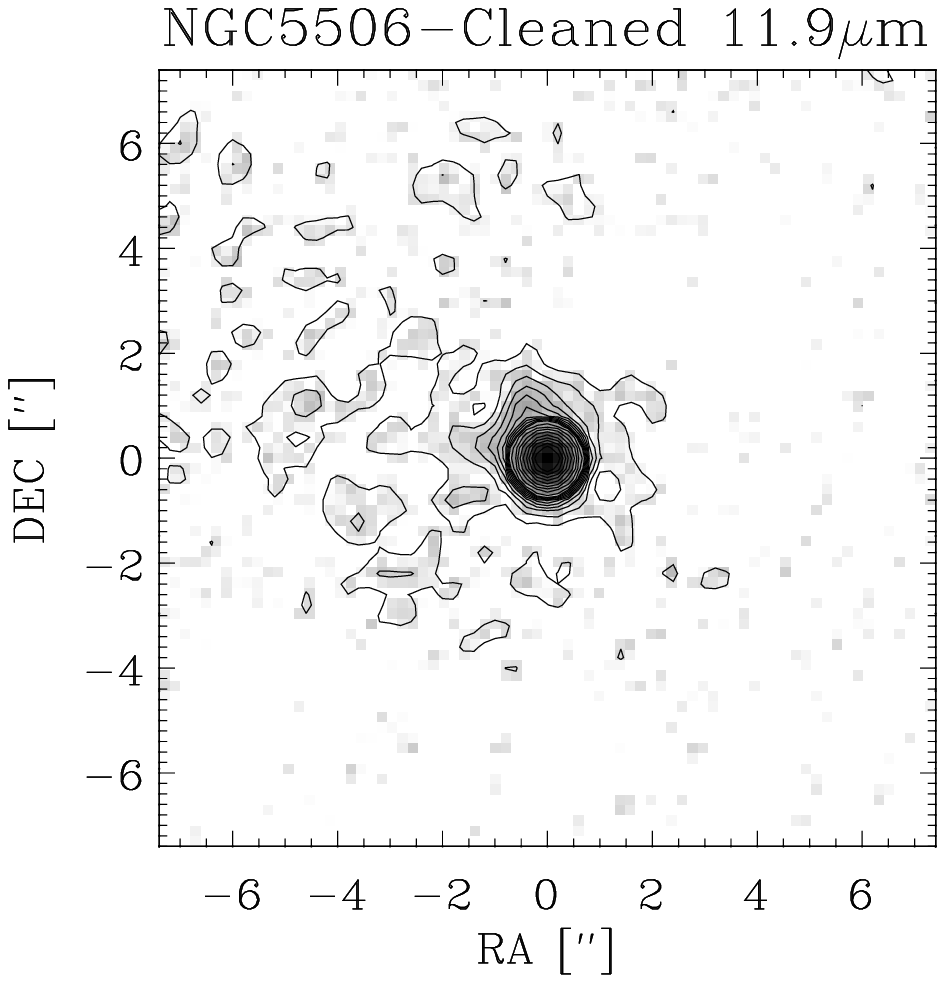}
 \includegraphics[width=7cm]{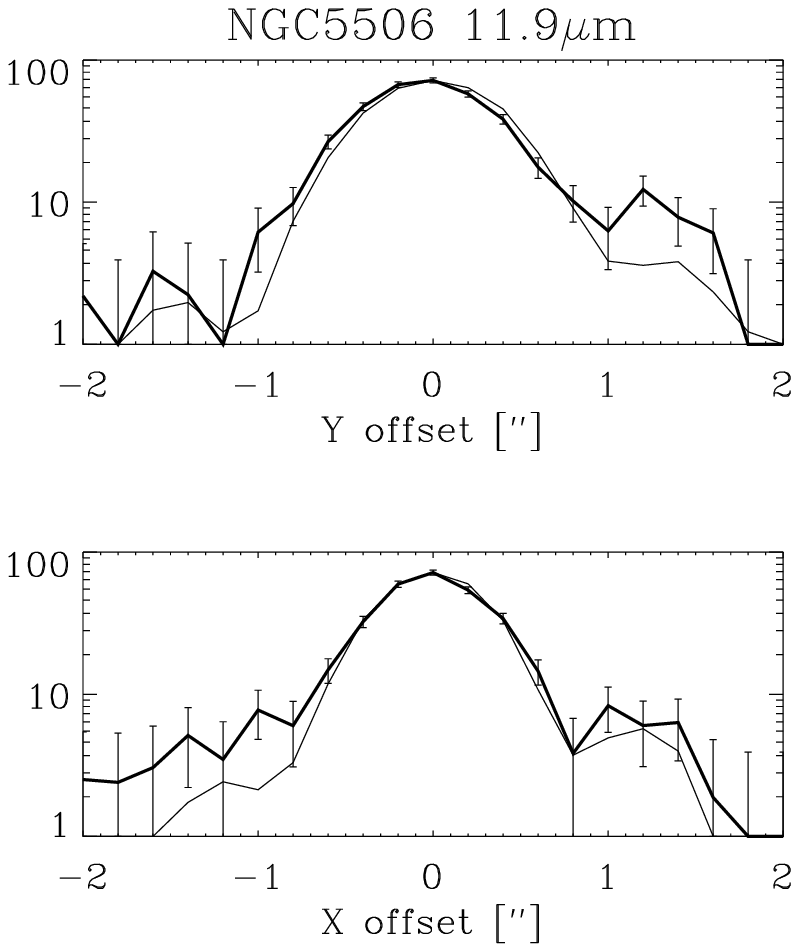}
  \caption{NGC 5506 contour and image overlays, see Figure
\ref{ngc1365}. \label{N5506}}
\end{figure*}

\begin{figure*}[htbp]
 \includegraphics[width=7cm]{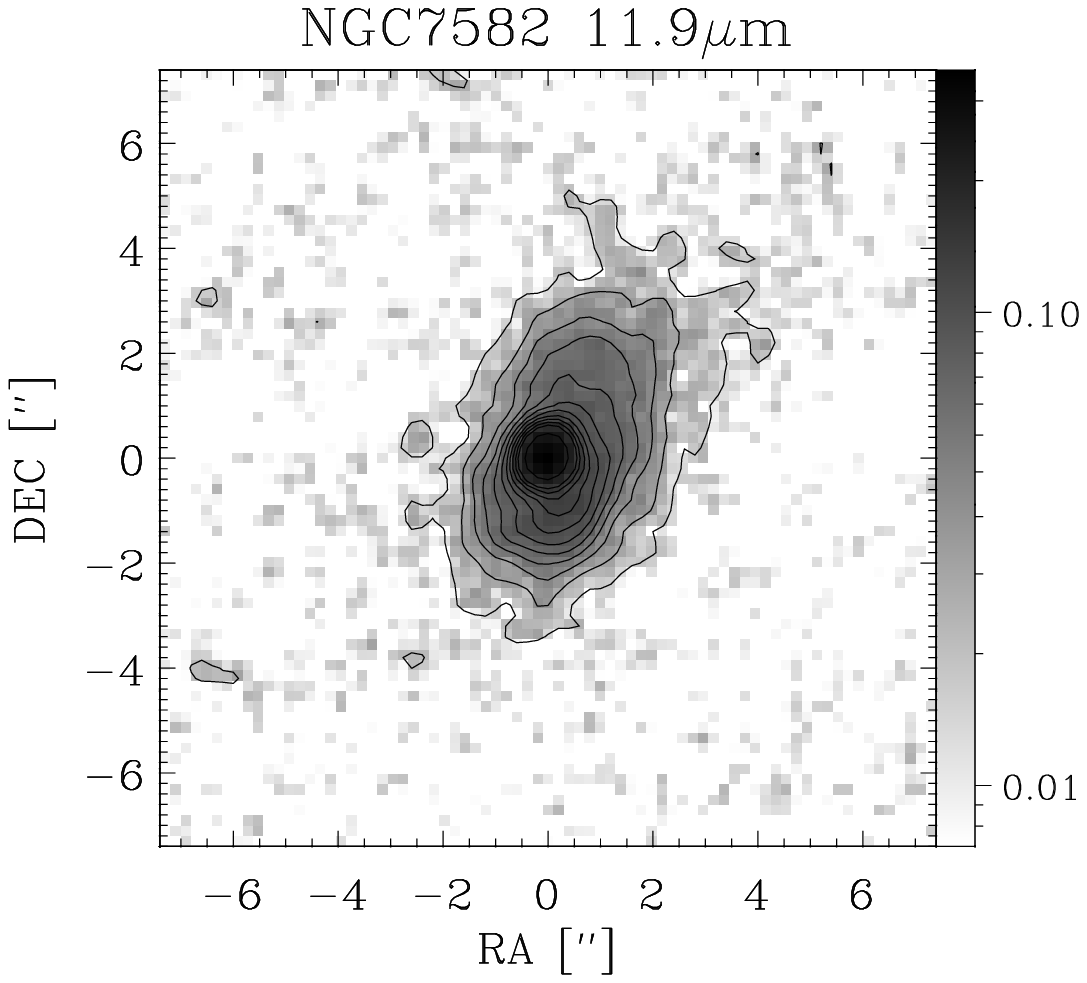}
 \includegraphics[width=7cm]{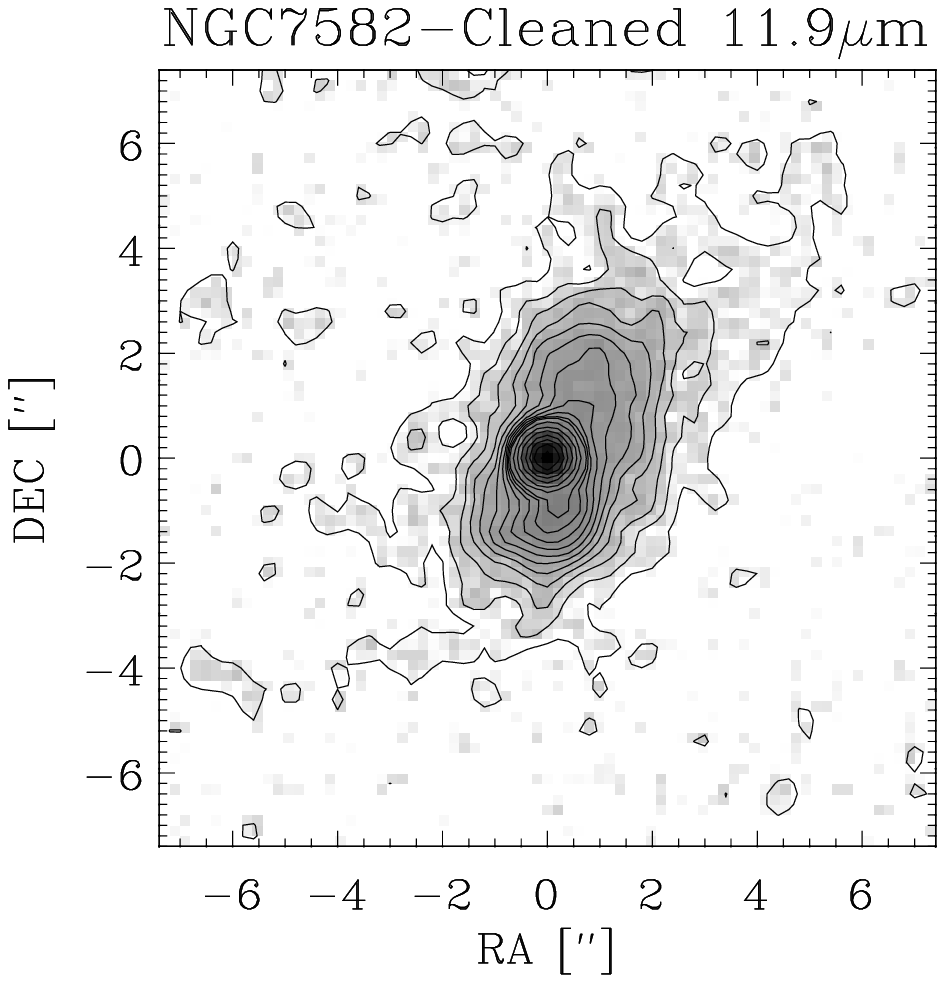}
  \includegraphics[width=7cm]{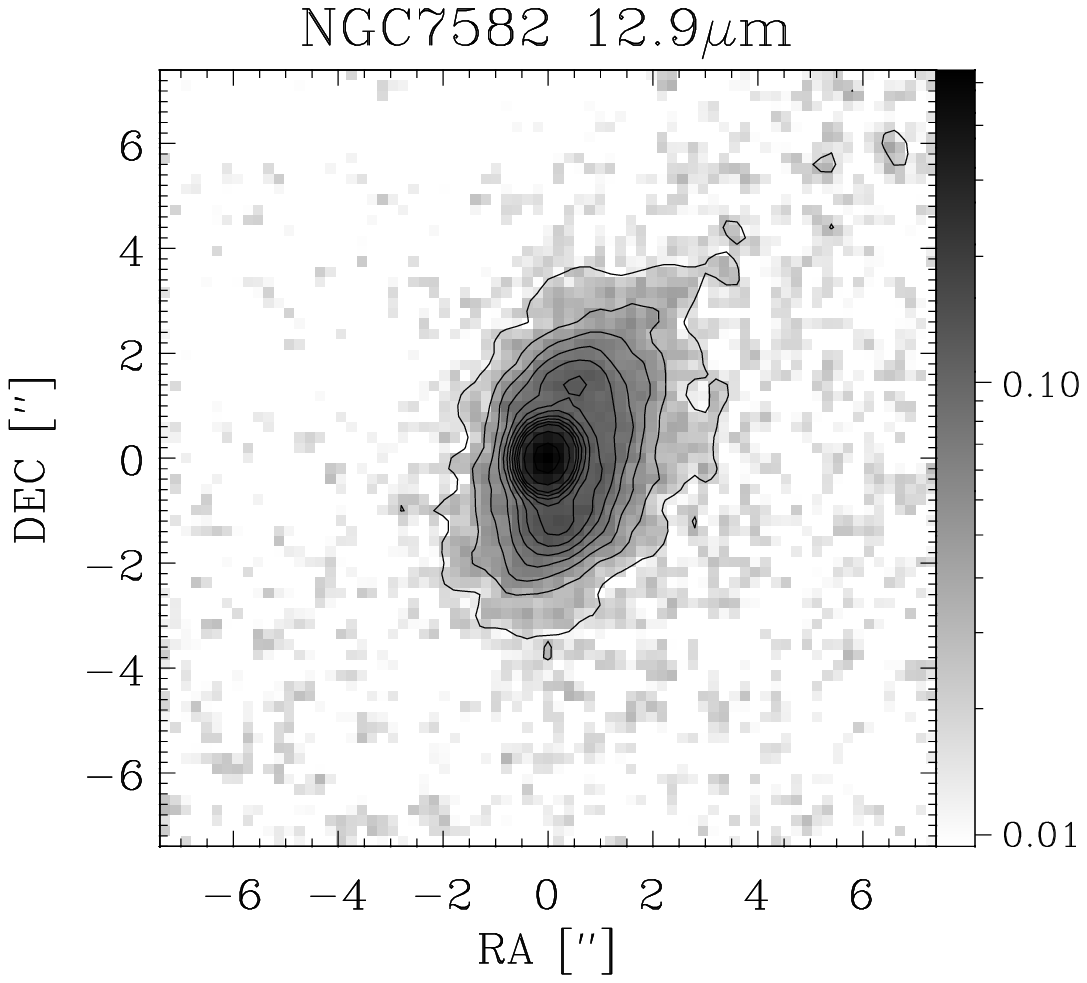}
  \includegraphics[width=7cm]{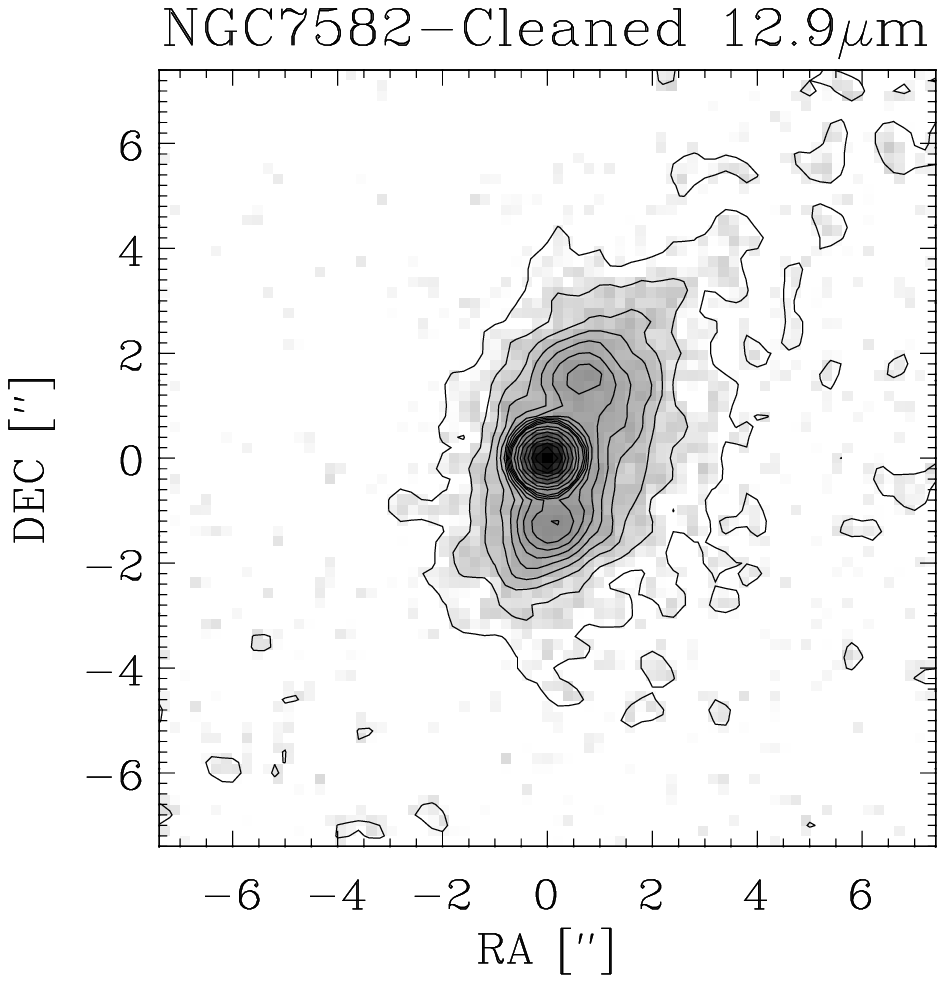}
\includegraphics[width=7cm]{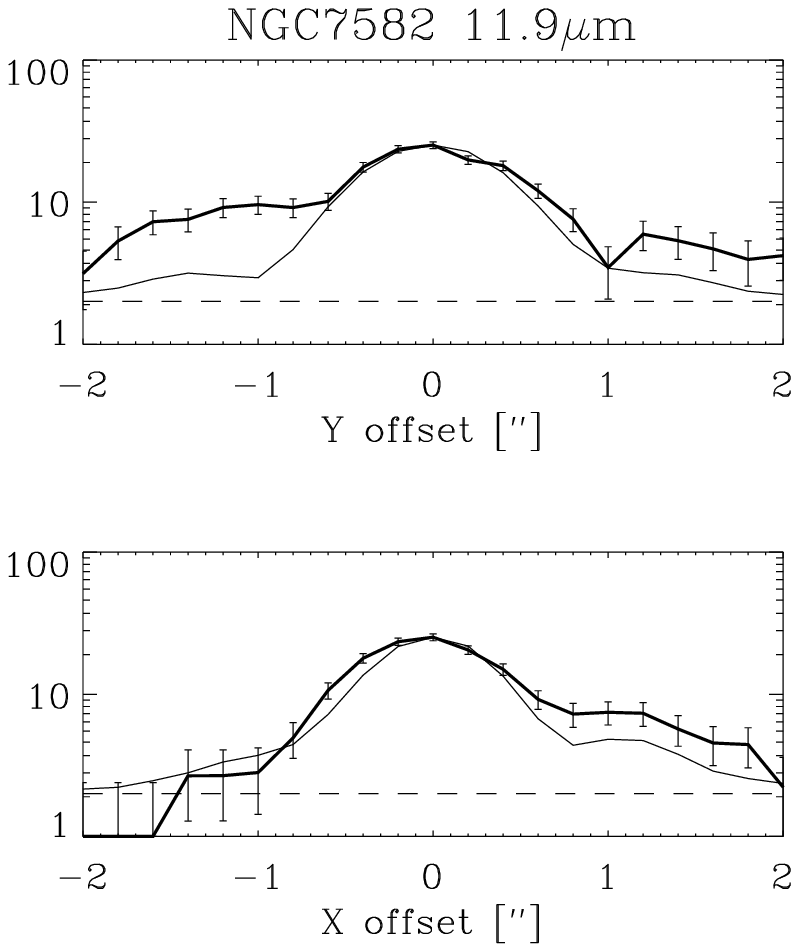}
\includegraphics[width=7cm]{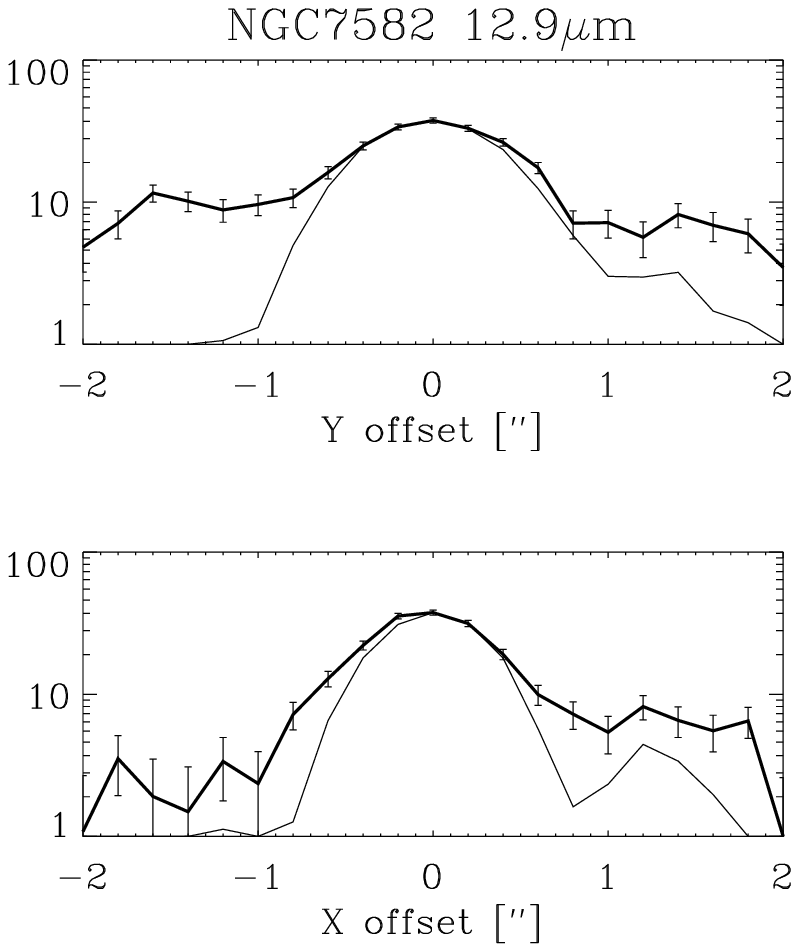}

  \caption{NGC 7582 contour and image overlays, see Figure
\ref{ngc1365}. \label{N7582}}
\end{figure*}

Core flux measurements and core positions for each source are presented in Table \ref{tab:1}.
Figures \ref{N253b} to \ref{N7582} show high resolution contour
images of eight objects out of the 21 objects observed, omitting
those sources for which  more recent images are available and images of unresolved sources.  A brief discussion of each source is
presented, along with references for other similar or complementary data. All sources show an unresolved core  with little or no
extended emission, with the exception of NGC 253, NGC 7582, NGC1365, and M83, which show considerable extended emission. 

\paragraph{NGC 253}
Figure \ref{N253b}. This nearby starburst galaxy shows complex extended
emission in all observed wavelengths: 8.6, 11.9 and 12.9\mum\,. All
images show
two sources, the bright source seems to be resolved. The extended
emission shows an elongation of 3'' to the north east. An additional
peak is
seen clearly in the 8.6 \mum\, image, and to a lesser extent in the 11.9\mum\, image, and corresponds to "peak 3" that \cite{Kalas94}
identify at 3.28\mum\, and is most likely to be PAH emission. 
The 12.9 \mum\ image, shows the structure boarded by the two peaks
reported by \cite{Boeker98} as well as other
authors, and is very similar to the $NeII$ map of these authors and
the $NeII$ maps of \cite{Keto99}. It is most likely to be
dominated by a combination of $NeII$ emission and 12.7\mum\, PAH
emission (\cite{Boeker98}, \cite{Forster03}). Since the 11.9
 filter is centered on 11.66\mum\, 
 with a FWHM of 1.16\mum\, it is likely that the 11.9\mum\, extended
emission also includes  11.3\mum\, PAH emission. \cite{Galliano05}
show 11.9\mum\, deconvolved images and identify six sources, as
opposed to four in our 11.9\mum\, image
Comparing the two images, sources M2 and M3 of \cite{Galliano05}
coincide with our second brightest peak, and sources M5 and M6  with
the two peaks to the North-East. The  flux measured by
\cite{Galliano05} for the main peak ($5000\pm 1000$mJy) is a factor of
two higher
than the flux measured in this work. This discrepancy can be attributed to the 2'' aperture
 used by \cite{Galliano05} as opposed to out method of measuring the peak value (see \S\ref{sec:reduction})
Identifying out second peak with M2+M3 our flux (1150 mJy) is  50 \% higher than the combined flux of M2 and M3.

\paragraph{NGC 1365} 
 Figure \ref{ngc1365}.   A face-on spiral galaxy with a
prominent bar. Our $11.9$\mum\, image shows an unresolved nucleus and
two point sources to
the north east, identified by \cite{Galliano05} as M5 and M6. This structure is surrounded
 by an arm like faint extended emission. In addition, the deconvolved image shows slight point
 like emission from sources M7 and M4 of \cite{Galliano05}. The authors identify their M4, M5 and M6
 sources with radio counterparts of \cite{Saikia94}, and conclude that these sources are embedded young massive
star clusters.  Flux measurements of the nucleus are in excellent agreement with \cite{Galliano05}, but are
 higher than those of  \cite{Siebenmorgen04} who measured a flux of  400mJy at 8.5\mum\, and 460mJy
 at 10.4\mum. This discrepancy can be explained by the multi-source structure of NGC 1365, causing flux
 measurements very sensitive to telescope positioning. The weak extended emission to the South
West of the nucleus also appears as two point sources in the 11.9 and 10.4 \mum\, deconvolved
maps of \cite{Galliano05} and is most likely not an artifact.

\paragraph{IRAS 05189-2524}
Observations of  this Seyfert 2 galaxy at $11.9 \mu m$ 
 show a completely unresolved core with no deviations from the PSF larger then $ 1 \sigma$. An
unresolved core is also reported by   \cite{Siebenmorgen04} with  8.6, 10.4 and 11.9\mum\, flux densities
of 420,420 and 570\,mJy, respectively, and by
   \cite{Soifer00}, who present a 12.5\mum\, image and  
12.5\,24.5\mum\, fluxes.

\paragraph{NGC 2377} For this object we give an upper limit for the 
flux of 60mJy.

\paragraph{MCG 5-23-16}
For this S0 galaxy hosting a Seyfert 1.9 nucleus we have an unresolved core with a flux of 646 mJy at 11.9\mum.

\paragraph{Mrk1239} Figure \ref{Mrk1239} Observations of this highly
polarised narrow-line Seyfert 1 galaxy at 11.9\mum\, shows central
source unresolved.
 The 11.9\mum\, flux is in agreement with \cite{Maiolino95}.

\paragraph{NGC 3256}
Our image (not presented here) of this IR-luminous merger system  shows an unresolved core, In contrast to 
\cite{Siebenmorgen04} who present  a resolved, yet featureless image, taken with shorter exposure time.
The difference may be explained by a change in seeing between the calibrator, from which the PSF
 is determined, and NGC 3256.

\paragraph{NGC 3281} For this Seyfert 2 galaxy we measure 625 mJy for 11.9 \mum, which is similar
 to the N-band flux found at  \cite{Krabbe01}   $580\pm30\,$mJy, measured with a 2.2m
telescope, and therefore a diffraction limit 1.6 times that of our observation.
 A $18''\times20''$ 10.5 \mum\, image
 of NGC 3281  dominated by a point source and marginal evidence for an extended emission component
is also presented by \cite{Krabbe01}. 

\paragraph{NGC 3758}(Mkn 739) For this double-nuclei Seyfert I/starburst galaxy  a low flux of $<60\,$mJy is measured.

\paragraph{NGC 3783} NGC 3783 is a nearly face-on SBa galaxy with a very
bright, highly variable, Seyfert 1 nucleus. The image shows an
unresolved point source.

\paragraph{3C273} The  image of this well known quasar shows an unresolved core at 11.9\mum. The measured 11.9\mum\, flux is in
 agreement with the N-band flux of \cite{Sitko82} but  100mJy higher than the flux quoted in \cite{Gorjian04}.

\paragraph{NGC 4594} "Sombrero Galaxy". For this object the measured
11.9\mum\, flux, $<60 mJy$, is in agreement with the results
 found at \cite{Maiolino95} and \cite{Gorjian04}.

\paragraph{MCG-3-34-6} The low flux measured here, $<60$\,mJy at
11.9\mum is much smaller than the 440 mJy flux measured by
\cite{Maiolino95} at 10.4\mum, with a $5.4''$ aperture. The origin of
the discrepancy are unclear since this object has not been studied
before apart from the single measurement of \cite{Maiolino95}
mentioned above, and so it is possible that the N-band emission is
variable. Our image shows an unresolved point source and therefore it
is unlikely that extended emission is responsible for the discrepancy.
 
\paragraph{NGC 5128} Figure \ref{N5128}  Centaurus A, the closest
active radio galaxy. The 11.9 \mum\, image shows an unresolved central
core, with extended  emission at 10\% level.
The 8.6\mum\, image shows an unresolved core. For comparison, see
\cite{Siebenmorgen04} for a 10.4\mum\, image showing an unresolved
 core of less than $0.5''$.  

\paragraph{M\,83}
Figure \ref{M83}
This near, face-on barred spiral galaxy shows mostly extended emission surrounding a faint central
 object. The extended emission is dominated by PAHs which account for the majority if the MIR
luminosity \citep{Vogler05}. The $11.9\mu m$\, flux measured, $232$\,mJy,
is  identical with the $11.9 \mu m$ \, flux measured by \cite{Siebenmorgen04}. Our images, however,
show a clear, although faint, central source, while \cite{Siebenmorgen04} report only fuzzy extended emission.
The LW3 filter ($12-18$\mum) of \cite{Vogler05} show the emission clearly tracing the spiral arms of M83, which
 is not seen in our image.

\paragraph{ESO 445-G50}
For this Seyfert I galaxy we measure 352 mJy at 11.9\mum. A 10.4 \mum\, flux of 640 mJy can
 be found at \cite{Siebenmorgen04}.

\paragraph{Mrk463} The image of this double-nuclei Seyfert 2 galaxy
 shows an unresolved core at 11.9\mum, measuring 338 mJy.

\paragraph{Circinus} Figure \ref{circinus} Observations of this nearby
spiral were made at 11.9 \mum and 8.6 \mum. The seeing at for the
11.9 \mum point was bad, so no statement can be made whether the central
 peak is resolved or not. The  8.6 \mum does show a slightly resolved background component yet
whether this is a disk or  circum-nuclear emission (reported by \cite{Krabbe01} with a radius
 of $\approx 1''$ is not clear. The image presented here is very similar to the one presented by
\cite{Siebenmorgen04}.
Recent high resolution 8.74 and 18.33\mum\, images and flux
measurements are also presented by \cite{Packham05}, which also find a
higher 8.74\mum\, flux, 5.5-8.4\,Jy, for a range of apertures from $1''$ to $5''$.

\paragraph{NGC 5506} Figure \ref{N5506} For this edge-one irregular
Seyfert 1 (S1i) galaxy ,we show here the first high resolution images
at 8.6 and 11.
9\mum. The 8.6\mum\, image shows an unresolved core while extended emission to the north-east is
 seen in the $11.9 \mu m$ image. The 11.9\mum\, flux measured, 908 mJy, is
similar to the \cite{Siebenmorgen04} measurement of 1060 mJy.

\paragraph{NGC 7469}
For this prototypical Seyfert 1 galaxy we encountered an unknown image quality problem.
 \cite{Soifer03} present a high resolution  ($4''\times4''$) 12.5 \mum \, image of the nucleus
 of NGC  7469, resolving the ring structure around the nucleus and an extended   structure in the nucleus itself.
\paragraph{NGC 7582}
NGC 7582 is a classic Seyfert 2 galaxy which, during a period of five months in 1999 showed broad lines
 characteristic of a Seyfert 1  galaxy, which may indicate the presence of a patchy torus. \cite{Aretxaga99}.
 50\% of the flux  found in extended emission with a peak offset
from the center of the extended emission.  The extended emission is seen elongated along the South-North direction.
The 12.9\mum\, image shows two weak sources to the North and to the South of the main peak, also clearly
 seen in the 12.9\mum\, image  of \cite{Acosta-Pulido03}. The 11.9\mum\, image shows one slightly resolved
 peak surrounded by extended emission. Flux at 11.9\mum\,   is very similar to that of \cite{Siebenmorgen04},
 although the 11.9\mum\, image they present shows considerably less extended    emission. Similar extended
emission can also be seen in the  N1 ($\approx 8-95.5$\mum) image of \cite{Acosta-Pulido03}.

\begin{table*}

 \caption{Observation list and fluxes for all objects.}
\label{tab:1} 
\centering
\begin{tabular}{c|cc|c|c|c|cc|c}

\hline\noalign{\smallskip}

Name & RA(J2000) & DEC(J2000)&  D [Mpc] &  \multicolumn{2}{|c|}{Flux [mJy]} & Date & Exp. [sec] &Apreture [''] \\
\cline{5-6}
     &           &        &      &             11.9$\mu$m & 8.9$\mu$m    &      &      &     \\
\hline
N $253^1$      & 00 47 33.1   & -25 17 17.2    & 3.3    &  2040-2800 & 1140 & 6 Aug. 2001          & 2955 &1.2\\
N $253^2$      &                    &                        &          &1160            & 695    &                              &           &1.2               \\
N $1365^1$     & 03 33 36.4  & -36 08 25.5    & 20.7    &  606       & -    & 6 Aug. 2001 & 5373 &1.2 \\
N $1365^2$     &                   &                     &            &  157       & -    &             &      & 1.2\\
N $1365^3$     &                   &                     &            &  152       & -    &             &      &1.2\\
IRS 05189-2524 & 05 21 01.4   & -25 21 44.9   & 172.6   &  545       & -    & 7 Aug. 2001 & 2148 & 1.6\\
N 2377              & 07 24 56.8  & -09 39 36.9   & 31.3      &  $<60$     & -    & 8 Feb. 2002 & 752  & 1.6\\
MCG-5-23-16     & 09 47 40.2  & -30 56 54.2   & 31.9      &  648       & -    & 8 Feb. 2002 & 2256  &1.6\\
Mrk 1239          & 09 52 19.1  &  -01 36 43.5  & 79.7      &  638       & -    & 8 Feb. 2002 & 1504  & 1.6\\
N 3256             & 10 27 51.8  & -43 54 08.7   & 35.4      &  553       & -    & 8 Feb. 2002 & 1504   &1.6\\
N 3281             & 10 31 52.0  & -34 51 13.3   & 40.9      &  625       & -    & 8 Feb. 2002 & 1880   &1.6\\
N 3758             & 11 36 29.0  & +21 35 47.8  & 122      &  $<60$     & -    & 8 Feb. 2002 & 752 & 1.6\\
N 3783             & 11 39 01.8  & -37 44 18.7   & 37.2      &  590       & -    & 7 Aug. 2001 & 1074 &1.6\\
3C 273             & 12 29 06.7  &  +02 03 08.5 & 649      &  345       & -    & 8 Feb. 2002 & 1504  &1.6\\
N 4594             & 12 39 59.4  & -11 37 23.0  & 12.4       &  $<60$     & -    & 7 Aug. 2001 & 537  &1.6\\
MCG-3-34-6      & 13 10 23.7  & -21 41 09.0  & 95.4       &  $<60$     & -    & 7 Aug. 2001 & 537  &1.6\\
N 5128             & 13 25 27.6  & -43 01 08.8  & 5.3         &  1220      & 635  & 6 Aug. 2001 & 2149 &1.6\\
M 83                & 13 37 00.8  & -29 51 58.6  & 5.2         &  232       & -    & 6 Aug. 2001 & 1209 &1.2\\
ESO 445-G50    & 13 49 19.3  & -30 18 34.4  & 64.2       &  352       & -    & 6 Aug. 2001 & 2149 &1.6\\
Mrk 463           & 13 56 02.9  & +18 22 19.5 & 20.7       &  338       & -    & 7 Aug. 2001 & 1611 &1.6\\
Circinus           & 14 13 09.3  & -65 20 20.6  & 3.6          &  9700      & 4700 & 6 Aug. 2001 & 1074& 1.6\\
N 5506            & 14 13 15.0  & -03 12 27.2  & 25           &  908       & -    & 7 Aug. 2001 & 1074 &1.6\\
N 7469            & 23 03 15.6  & +08 52 26.4 & 69.2   &  414       & -    & 6 Aug. 2001 & 2687 &1.6\\
N 7582            & 23 18 23.5  & -42 22 14.0   &  21.3   &  670       & -    & 6.Aug.2001  & 2687 &2.6\\
\noalign{\smallskip}\hline
\end{tabular}
\footnote Table notes. Ra and Dec coordinates are core coordinates taken  directly from the telescope
position. Distances are from NED.  RA \& DEC error: $\pm5''$. Flux errors are $\pm 15\%$.
\end{table*}

\section{Conclusion}\label{sec:conclusion}

We presented new high resolution mid-infrared images and fluxes at 8.9,11.9 and 12.9\mum\, for the
brightest AGN's observable from Cerro-Paranal. Most  sources show an unresolved core, with little or no
extended emission.  Considerable extended emission has only been detected for
 NGC 253, NGC 7582 and M83. For each source, a brief discussion is given along with a comparison
 with other relevant published data. In general we find our images and fluxes to be in agreement
with previous papers.



\bibliography{bib}

\begin{thebibliography}{25}
\expandafter\ifx\csname natexlab\endcsname\relax\def\natexlab#1{#1}\fi

\bibitem[{{Acosta-Pulido} {et~al.}(2003){Acosta-Pulido}, {P{\' e}rez
  Garc{\'{\i}}a}, {Prieto}, {Rodr{\'{\i}}guez-Espinosa}, \& {Cair{\'
  o}s}}]{Acosta-Pulido03}
{Acosta-Pulido}, J.~A., {P{\' e}rez Garc{\'{\i}}a}, A.~M., {Prieto}, M.~A.,
  {Rodr{\'{\i}}guez-Espinosa}, J.~M., \& {Cair{\' o}s}, L.~M. 2003, in Revista
  Mexicana de Astronomia y Astrofisica Conference Series, 198--201

\bibitem[{{Aretxaga} {et~al.}(1999){Aretxaga}, {Joguet}, {Kunth}, {Melnick}, \&
  {Terlevich}}]{Aretxaga99}
{Aretxaga}, I., {Joguet}, B., {Kunth}, D., {Melnick}, J., \& {Terlevich}, R.~J.
  1999, \apjl, 519, 123

\bibitem[{{Boeker} {et~al.}(1998){Boeker}, {Krabbe}, \& {Storey}}]{Boeker98}
{Boeker}, T., {Krabbe}, A., \& {Storey}, J.~W.~V. 1998, \apjl, 498, L115+

\bibitem[{{F{\"o}rster Schreiber} {et~al.}(2003){F{\"o}rster Schreiber},
  {Sauvage}, {Charmandaris}, {Laurent}, {Gallais}, {Mirabel}, \&
  {Vigroux}}]{Forster03}
{F{\"o}rster Schreiber}, N.~M., {Sauvage}, M., {Charmandaris}, V., {et~al.}
  2003, \aap, 399, 833

\bibitem[{{Galliano} {et~al.}(2005){Galliano}, {Alloin}, {Pantin}, {Lagage}, \&
  {Marco}}]{Galliano05}
{Galliano}, E., {Alloin}, D., {Pantin}, E., {Lagage}, P.~O., \& {Marco}, O.
  2005, \aap, 438, 803

\bibitem[{{Gorjian} {et~al.}(2004){Gorjian}, {Werner}, {Jarrett}, {Cole}, \&
  {Ressler}}]{Gorjian04}
{Gorjian}, V., {Werner}, M.~W., {Jarrett}, T.~H., {Cole}, D.~M., \& {Ressler},
  M.~E. 2004, \apj, 605, 156

\bibitem[{{Granato} {et~al.}(1997){Granato}, {Danese}, \&
  {Franceschini}}]{Granato97}
{Granato}, G.~L., {Danese}, L., \& {Franceschini}, A. 1997, \apj, 486, 147

\bibitem[{{K{\" a}ufl} {et~al.}(2003){K{\" a}ufl}, {Sterzik}, {Siebenmorgen},
  {Weilenmann}, {Relke}, {Hron}, \& {Sperl}}]{Kaufl03}
{K{\" a}ufl}, H., {Sterzik}, M.~F., {Siebenmorgen}, R., {et~al.} 2003, in
  Instrument Design and Performance for Optical/Infrared Ground-based
  Telescopes. Edited by Iye, Masanori; Moorwood, Alan F. M. Proceedings of the
  SPIE, Volume 4841, pp. 117-128 (2003)., 117--128

\bibitem[{{Kalas} \& {Wynn-Williams}(1994)}]{Kalas94}
{Kalas}, P. \& {Wynn-Williams}, C.~G. 1994, \apj, 434, 546

\bibitem[{{Keto} {et~al.}(1999){Keto}, {Hora}, {Fazio}, {Hoffmann}, \&
  {Deutsch}}]{Keto99}
{Keto}, E., {Hora}, J.~L., {Fazio}, G.~G., {Hoffmann}, W., \& {Deutsch}, L.
  1999, \apj, 518, 183

\bibitem[{{Krabbe} {et~al.}(2001){Krabbe}, {B{\" o}ker}, \&
  {Maiolino}}]{Krabbe01}
{Krabbe}, A., {B{\" o}ker}, T., \& {Maiolino}, R. 2001, \apj, 557, 626

\bibitem[{{Leinert} {et~al.}(2004){Leinert}, {van Boekel}, {Waters},
  {Chesneau}, {Malbet}, {K{\"o}hler}, {Jaffe}, {Ratzka}, {Dutrey}, {Preibisch},
  {Graser}, {Bakker}, {Chagnon}, {Cotton}, {Dominik}, {Dullemond},
  {Glazenborg-Kluttig}, {Glindemann}, {Henning}, {Hofmann}, {de Jong},
  {Lenzen}, {Ligori}, {Lopez}, {Meisner}, {Morel}, {Paresce}, {Pel},
  {Percheron}, {Perrin}, {Przygodda}, {Richichi}, {Sch{\"o}ller}, {Schuller},
  {Stecklum}, {van den Ancker}, {von der L{\"u}he}, \& {Weigelt}}]{Leinert04}
{Leinert}, C., {van Boekel}, R., {Waters}, L.~B.~F.~M., {et~al.} 2004, \aap,
  423, 537

\bibitem[{{Lipovetsky} {et~al.}(1988){Lipovetsky}, {Neizvestny}, \&
  {Neizvestnaya}}]{Lipovetsky88}
{Lipovetsky}, V.~A., {Neizvestny}, S.~I., \& {Neizvestnaya}, O.~M. 1988,
  Soobshcheniya Spetsial'noj Astrofizicheskoj Observatorii, 55, 5

\bibitem[{{Maiolino} {et~al.}(1995){Maiolino}, {Ruiz}, {Rieke}, \&
  {Keller}}]{Maiolino95}
{Maiolino}, R., {Ruiz}, M., {Rieke}, G.~H., \& {Keller}, L.~D. 1995, \apj, 446,
  561

\bibitem[{{Meisenheimer} {et~al.}(2007){Meisenheimer}, {Tristram}, {Jaffe},
  {Israel}, {Neumayer}, {Raban}, {R{\"o}ttgering}, {Cotton}, {Graser},
  {Henning}, {Leinert}, {Lopez}, {Perrin}, \& {Prieto}}]{Meisenheimer07}
{Meisenheimer}, K., {Tristram}, K.~R.~W., {Jaffe}, W., {et~al.} 2007, \aap,
  471, 453

\bibitem[{{Ohnaka} {et~al.}(2005){Ohnaka}, {Bergeat}, {Driebe}, {Graser},
  {Hofmann}, {K{\"o}hler}, {Leinert}, {Lopez}, {Malbet}, {Morel}, {Paresce},
  {Perrin}, {Preibisch}, {Richichi}, {Schertl}, {Sch{\"o}ller}, {Sol},
  {Weigelt}, \& {Wittkowski}}]{Ohnaka05}
{Ohnaka}, K., {Bergeat}, J., {Driebe}, T., {et~al.} 2005, \aap, 429, 1057

\bibitem[{{Reimann} {et~al.}(2000){Reimann}, {Linz}, {Wagner}, {Relke},
  {Kaeufl}, {Dietzsch}, {Sperl}, \& {Hron}}]{Reimann00}
{Reimann}, H., {Linz}, H., {Wagner}, R., {et~al.} 2000, in Proc. SPIE Vol.
  4008, p. 1132-1143, Optical and IR Telescope Instrumentation and Detectors,
  Masanori Iye; Alan F. Moorwood; Eds., 1132--1143

\bibitem[{{Saikia} {et~al.}(1994){Saikia}, {Pedlar}, {Unger}, \&
  {Axon}}]{Saikia94}
{Saikia}, D.~J., {Pedlar}, A., {Unger}, S.~W., \& {Axon}, D.~J. 1994, \mnras,
  270, 46

\bibitem[{{Siebenmorgen} {et~al.}(2004){Siebenmorgen}, {Kr{\" u}gel}, \&
  {Spoon}}]{Siebenmorgen04}
{Siebenmorgen}, R., {Kr{\" u}gel}, E., \& {Spoon}, H.~W.~W. 2004, \aap, 414,
  123

\bibitem[{{Sitko} {et~al.}(1982){Sitko}, {Stein}, {Zhang}, \&
  {Wisniewski}}]{Sitko82}
{Sitko}, M.~L., {Stein}, W.~A., {Zhang}, Y.-X., \& {Wisniewski}, W.~Z. 1982,
  \apj, 259, 486

\bibitem[{{Soifer} {et~al.}(2003){Soifer}, {Bock}, {Marsh}, {Neugebauer},
  {Matthews}, {Egami}, \& {Armus}}]{Soifer03}
{Soifer}, B.~T., {Bock}, J.~J., {Marsh}, K., {et~al.} 2003, \aj, 126, 143

\bibitem[{{Soifer} {et~al.}(2000){Soifer}, {Neugebauer}, {Matthews}, {Egami},
  {Becklin}, {Weinberger}, {Ressler}, {Werner}, {Evans}, {Scoville}, {Surace},
  \& {Condon}}]{Soifer00}
{Soifer}, B.~T., {Neugebauer}, G., {Matthews}, K., {et~al.} 2000, \aj, 119, 509

\bibitem[{{Starck} {et~al.}(1999){Starck}, {Abergel}, {Aussel}, {Sauvage},
  {Gastaud}, {Claret}, {Desert}, {Delattre}, \& {Pantin}}]{Starck99}
{Starck}, J.~L., {Abergel}, A., {Aussel}, H., {et~al.} 1999, \aaps, 134, 135

\bibitem[{{Tristram} {et~al.}(2007){Tristram}, {Meisenheimer}, {Jaffe},
  {Schartmann}, {Rix}, {Leinert}, {Morel}, {Wittkowski}, {R{\"o}ttgering},
  {Perrin}, {Lopez}, {Raban}, {Cotton}, {Graser}, {Paresce}, \&
  {Henning}}]{Tristram07}
{Tristram}, K.~R.~W., {Meisenheimer}, K., {Jaffe}, W., {et~al.} 2007, \aap,
  474, 837

\bibitem[{{Vogler} {et~al.}(2005){Vogler}, {Madden}, {Beck}, {Lundgren},
  {Sauvage}, {Vigroux}, \& {Ehle}}]{Vogler05}
{Vogler}, A., {Madden}, S.~C., {Beck}, R., {et~al.} 2005, \aap, 441, 491

\end{thebibliography}
\Online
\begin{appendix}
\section{Figures}
\clearpage

\end{appendix}
\end{document}